\documentclass[12pt,a4paper]{article}
\usepackage{amsmath,amsthm,amsfonts,amssymb,alltt}
\usepackage{epsfig}
\usepackage{pdflscape}

\theoremstyle{definition}

\theoremstyle{remark}

\numberwithin{equation}{section}
\numberwithin{table}{section}
\numberwithin{figure}{section}

\usepackage{framed}

\usepackage{float}

\usepackage[section]{placeins}%

\begin{document}

\author{Wojciech Nawalaniec \\
Faculty of Mathematics, Physics and Technical Science\\
Pedagogical University of Cracow,\\ ul.~Podchorazych 2, Krakow 30-084,\\Poland}
\title{Classifying and analysis of random composites using structural sums feature vector
}
\date{}
\maketitle

\begin{abstract}
  The main goal of this paper is to present the application of structural sums, mathematical objects originating from the computational materials science, in construction of a feature space vector of 2D random composites simulated by distributions of non-overlapping disks on the plane.
  Construction of the feature vector enables the immediate application of machine learning tools and data analysis techniques to random structures.
  In order to present the accuracy and the potential of structural sums as geometry descriptors, we apply them to classification problems comprising composites with circular inclusions as well as composites with shapes formed by disks.  
    As an application, we perform the analysis of different models of composites in order to formulate the irregularity measure of random structures.  We also visualize the relationship between the effective conductivity of 2D composites and the geometry of inclusions.
\end{abstract}


\section{Introduction}\label{sect:intro}

The present paper is devoted to the study of {\it structural sums}, mathematical objects credited to Mityushev~\cite{RveMit2006}, as geometric features of 2D composites. Modern computational theory of random composites~\cite{GluMitNaw2017Book} is based on the structural sums introduced for 2D composites with circular non-overlapping inclusions. This leads to the precise geometrical description of heterogeneous media, in particular, to a new conception of the representative volume element. Moreover, the analysis can be extended to other shapes by covering the considered domain by an appropriate cluster of disks.

 In general, one can consider two-phase structures or images representing composites, when the phases are measurable plane sets~\cite{Adler2007}. Any measurable set on the plane can be approximated by a set non-overlapping disks. We do not discuss here the question of such approximations and assume that the preprocessing stage of the extraction of centres of disks and their radii is accomplished. Having the representation of the composite in form of disks, one can study the distribution of inclusions. For example, how do we show that the distributions differs, whether models fall into the same class, i.e. do inclusions have the same properties of their plane distributions? Can we model a certain distribution of inclusions? If so, how accurate is the model? Assume that we obtained a series of data representing perturbations of inclusions, how can this process be visualized? In fact, there is the information characterizing the distribution encoded in the data, but what kind of tool should be used in order to extract it? In order to constructively study the stated questions, it is sufficient to
 construct a proper data representation in form of a set of numeric parameters describing geometry of the composite. Such a representation is called {\it feature vector} in the machine learning literature. Construction of the feature vector enables the immediate application of machine learning tools and data analysis techniques to random structures.

The present paper describes an approach of constructing a feature vector based on structural sums.  The theory of structural sums is relatively new and its advantage has not been explicitly described yet. The idea of application of structural sums in extracting features from images was first proposed in the PhD thesis~\cite{NawPhD}. Since structural sums need extensive investigations of their properties, we have to treat them in the context of a black box. Hence, in order to demonstrate the role of structural sums as a carrier of geometric information, we consider theoretical classification problems and observe the accuracy of the feature vector in application to unknown samples.
We believe that investigating various abstract or real-world classification and regression problems may lead to a much better understanding of structural sums and their application to analysis of composites.

It is worth noting the existing applications of structural sums in analysis of different kind of data.
For example, Mityushev and Nawalaniec~\cite{MitNaw2015} used structural sums in the systematic investigation of dynamically changing microstructures.
Other applications cover, for instance, an analysis of collective behaviour of bacteria~\cite{CzaMit2017}, a description of random non-overlapping walks of disks on the plane~\cite{Naw2}, and parameterized comparison of composite materials obtained in different technological process~\cite{Kurtyka2017}. All the above results are based on the observation of differences between particular sums of low orders for identical disks and do not study structural sums as a tool, nor explore their geometrical meaning. On the other hand, the current paper focuses on general approach to the use of structural sums in the role of features of random polydispersed composites. We define the generic form of {\it structural sums feature vector} and clearly show that the vector is indeed a carrier of information. Moreover, we present examples demonstrating that an increase in the approximation of structural sums feature vector leads to the prediction accuracy gain in considered problems, i.e. sums of higher orders carry more information than the lower-order parameters. It is also shown that different kind of structures may require the application of a specific modification of the feature vector. Then we proceed with applications to random structures and composites, and investigate the geometrical meaning of the selected sums.

The present paper is organized as follows. 
In section~\ref{sect:theory}, we review the background material on structural sums.
In section~\ref{sect:theory}, we briefly discuss arguments for using structural sums in geometry analysis. In section~\ref{sect:feature_vector}, we propose the construction of the infinite structural sums feature vector and its approximations.
Section~\ref{sect:examples} covers examples of classification of simulated composites using consecutive approximations of the structural sums feature vector. The examples concern inclusions in form of randomly distributed disks (subsection~\ref{sect:disks}) and clusters of disks (subsection~\ref{sect:shapes}). Moreover, we perform the analysis of confusion matrices corresponding to classification problems.
In section~\ref{sect:applications}, we perform the visualization and the analysis of the {\it irregularity} of random structures. We also demonstrate how the geometry of inclusions reflects in the effective conductivity of considered models.
\newpage

\section{Structural sums: theory, geometry and computations}
\label{sect:theory}
\subsection{Background: Effective conductivity}
\label{sect:conductivity}

\begin{figure}[!h]
\centering
\includegraphics[clip, trim=40mm 35mm 40mm 30mm, width=.7\textwidth]{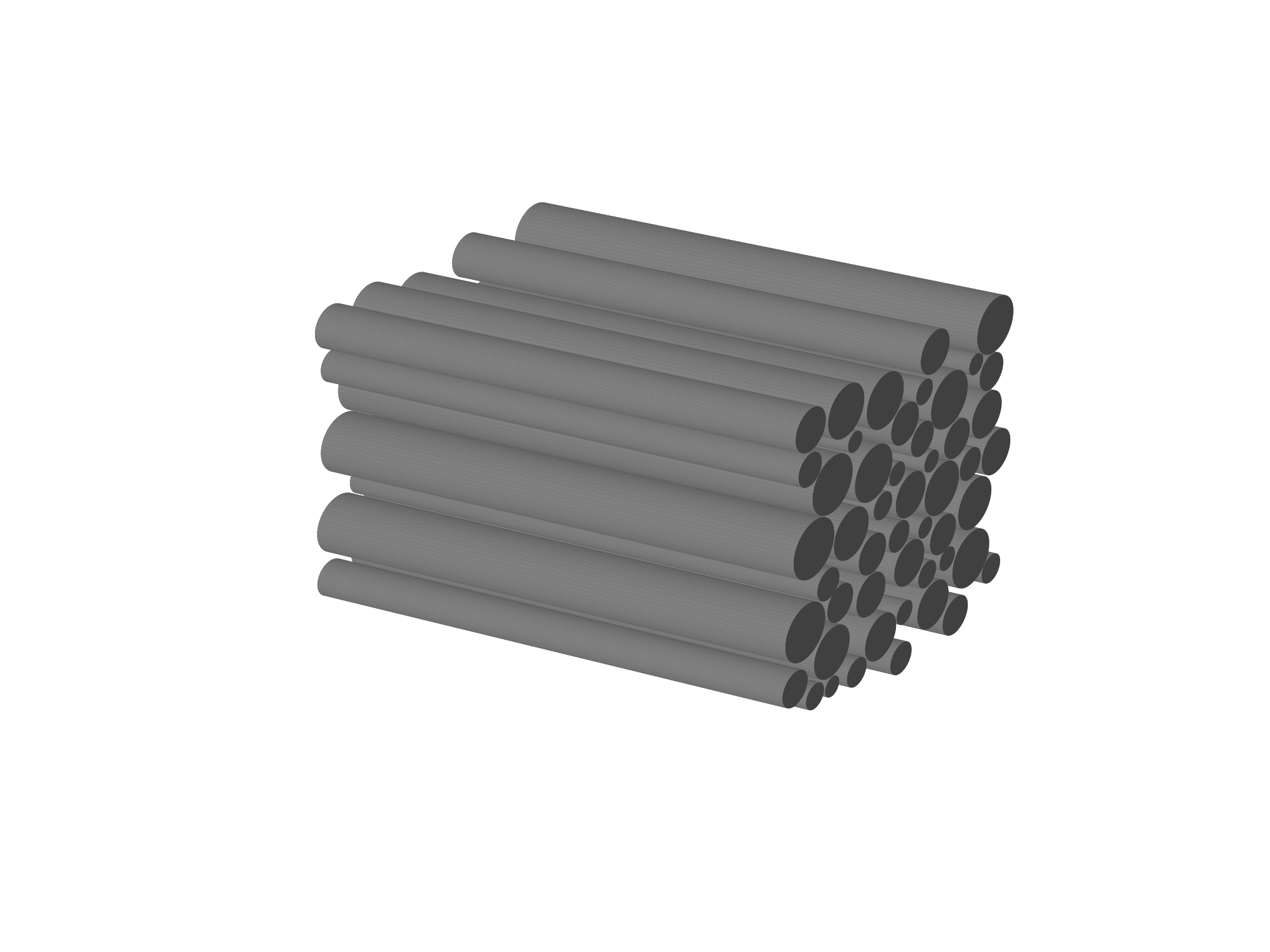} 
\caption{A model of polydispersed fibrous composite.} 
\label{fig:fibrous}
\end{figure}

\noindent Consider the effective conductivity of polydispersed fiber inclusions of conductivity $\lambda_f$ of different sizes randomly embedded in a matrix of conductivity normalized to unity, see Fig.~\ref{fig:fibrous}. 
A cross--section of such a composite is considered to be the doubly periodic two--dimensional lattice, see Fig.~\ref{fig:cell}. 
The effective conductivity of macroscopically isotropic 2D composites has the form~\cite{Mit2001}:
\begin{align*}
\lambda=1 +2\rho \nu \sum_{q=0}^{\infty }B_{q}\nu^{q},
\end{align*}
where $\rho=({\lambda_f-1})/({\lambda_f+1})$
 is the Bergman's contrast parameter~\cite{Bergman} and the constants $B_q$ are given as linear combinations of structural sums defined in section~\ref{sect:definition}. 
The algorithm for generating the symbolic representations of the coefficients $B_q$ takes the form of a recurrence relation of the first order~\cite{NawSymb2016}:
\begin{align*}
  \begin{array}{lll}
B_0=1, \; B_1=\pi^{-1}\rho e_{2}, \; B_2=\pi^{-2}\rho^2 e_{2,2},    \\ 
B_q=\pi^{-1}\beta B_{q-1}, \; q=3,4,5...,
\end{array}
\end{align*}
where $\beta$ is the {\it substitution operator} modifying every structural sum in $B_{q-1}$ according the transformation rule:
\begin{align*}
e_{p_1,p_2,\ldots,p_n}\longmapsto\rho e_{2,p_1,p_2,\ldots,p_n}-\frac{p_2}{p_1-1}e_{p_1+1,p_2+1,p_3,\ldots,p_n}.
\end{align*}


\subsection{Definition of structural sum}
\label{sect:definition}

\begin{figure}[!t]
\centering
\includegraphics[clip, trim=30mm 10mm 30mm 10mm, width=.7\textwidth]{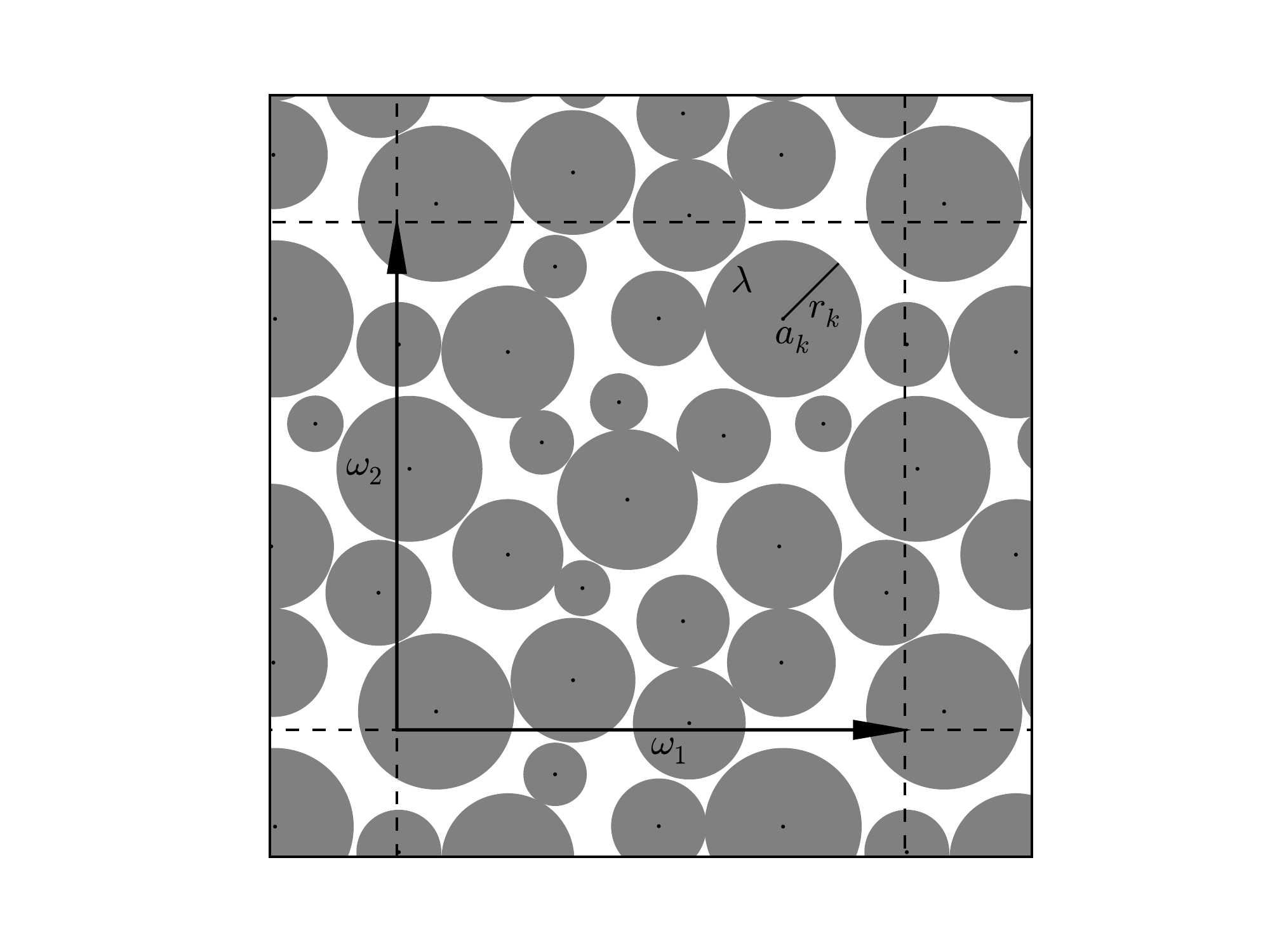} 
\caption{Doubly periodic cell $Q_{(0,0)}$ with a configuration of non-overlapping disks.} 
\label{fig:cell}
\end{figure}

\noindent Consider a periodic two--dimensional lattice $\mathcal{Q}$, defined by complex numbers $\omega_{1}$ and $\omega _{2}$ on the complex plane $\mathbb{C}$. The $(0,0)$-cell is introduced as the unit parallelogram  $Q_{(0,0)}:=\{z=t_{1}\omega_{1}+t_{2}\omega_{2}:-1/2<t_{j}<1/2\;(j=1,2)\}$. The lattice $\mathcal{Q}$ consists of the cells $Q_{(m_{1},m_{2})}:=\{z\in \mathbb{C}:z-m_{1}\omega_{1}-m_{2}\omega _{2}\in Q_{(0,0)}\}$, where $m_{1}$ and $m_{2}$ run over integer numbers. Consider $N$ non-overlapping disks of radii $r_j$ ($j=1,2,\ldots,N$) distributed in the $(0,0)$-cell ~(see~Fig.~\ref{fig:cell}). The total concentration of disks equals
\[ \nu=\pi\sum_{j=1}^N r_j^2. \]
Let $r$ be the largest of the radii $r_j$ ($j=1,2,\ldots,N$) 
and introduce constants $ \nu_j =\left(r_j / r\right)^2$ ($j=1,2,\ldots,N$) describing polydispersity, i.e. heterogeneity of sizes of disks.

Consider a set of points $a_k$ $(k= 1, 2,\ldots, N)$ being the centres of the disks. Let $n$ be a~natural number; $k_0, k_1\ldots, k_n$ be integers from 1 to
$N$; $k_j\geq 2$. Let $\mathbf{C}$ be the operator of the complex conjugation. The following sums were introduced by Mityushev~\cite{RveMit2006}:
\begin{equation}
	\label{eq:eSum}
	\begin{array}{c}
e^{\nu_0,\nu_1,\ldots,\nu_n}_{p_1,p_2,\ldots,p_n}=   \displaystyle{\frac{1}{\eta^{\delta+1}} {\sum_{k_0,k_1,\ldots,k_n}}} \nu^{t_0}_{k_0}\nu^{t_1}_{k_1}\nu^{t_2}_{k_2}\cdots\nu^{t_n}_{k_n} E_{p_1}(a_{k_0}-a_{k_1})\\ 
  \;\times\overline{E_{p_2}(a_{k_1}-a_{k_2})} E_{p_3}(a_{k_2}-a_{k_3})\cdots \mathbf{C}^{n+1} E_{p_n}(a_{k_{n-1}}-a_{k_n}),
	\end{array}
\end{equation}
\noindent where $\eta=\sum_{j=1}^N\nu_j$ and $\delta ={\frac{1}{2}\sum^n_{j=1}p_j}$. Functions $E_k$ ($k=2,3,\ldots$) are Eisenstein functions corresponding to the doubly periodic cell~$Q_{(0,0)}$ (see Appendix~\ref{eisenFun}), and the superscripts $t_j$ $(j=0,1,\ldots,n)$ are given by recurrence relations
\begin{equation}
  \label{eq:ep2nuk}
\begin{array}{ll}
t_0=1, \\
t_j=p_j-t_{j-1},\quad & j=1,2,\dots,n.
\end{array}
\end{equation}

\noindent The sum~\eqref{eq:eSum} is called the {\it structural sum of the multi-order} ${\mathbf p}=(p_{1},\ldots,p_{n})$. We also call $\delta$ the {\it order} of the structural sum. From this point, the superscripts for structural sums are omitted for the purpose of conciseness. In addition, following~\cite{NawSymb2016} we have $t_n=1$.

For example, in case of the composite modelled by $N$ identical disks, where $\nu_j=1$ ($j=1,2,\ldots,N$), structural sums $e_2$ and $e_{2,2}$ take the following forms:
\begin{equation*}
	\begin{array}{lll}
e_{2}= & \displaystyle{\frac{1}{N^2}} \displaystyle{\sum_{k_0=1}^{N}}\;\displaystyle{\sum_{k_1=1}^{N}}E_{2}(a_{k_0}-a_{k_1}),\\ \\
e_{2,2}= & \displaystyle{\frac{1}{N^3}} \displaystyle{\sum_{k_0=1}^{N}}\;\displaystyle{\sum_{k_1=1}^{N}}\;\displaystyle{\sum_{k_2=1}^{N}}E_{2}(a_{k_0}-a_{k_1})\overline{E_{2}(a_{k_1}-a_{k_2})}.
	\end{array}
\end{equation*}

\subsection{Structural sums and geometry}
\label{sect:geometry}

Let the properties of the composite be fixed. Then, the fundamental problem of composites consists in the construction of a homogenization operator $\mathcal H: G \to M$, where $G$
stands for microstructure (geometry) and $M$ for the macroscopic
physical constants. The key point is a precise and
convenient description of the geometrical set $G$ which can be
given, for instance, as a set of images. The recent research by Mityushev and Nawalaniec~\cite{MitNaw2015}, where structural sums were applied in the systematic investigation of the dynamically changing structures, proposes a choice of the geometric parameters as the following set of structural sums:
$$G =\{ e_{\mathbf m}, \;\mathbf m \in \mathcal M_e\},$$ 
where the set $\mathcal M_e$ is introduced by
the recursive rules~\cite{NawSymb2016}:
\begin{equation} \label{eq:2.18}
\begin{split}
   1.\quad & (2), (2,2)\in {\mathcal M_e};    \\
   2.\quad & \mbox{If} \ (m_{1},\ldots,m_{q})\in{\mathcal M_e} \ \mbox{and}\ q>1, \ \mbox{then} \\
    &(2,m_{1},\ldots,m_{q}),(m_{1}+1,m_{2}+1,m_{3},\ldots,m_{q})\in{\mathcal M_e}. 
\end{split}
\end{equation}

\begin{figure}[!t]
\centering
\includegraphics[clip, trim=25mm 0mm 15mm 5mm, width=.44\textwidth]{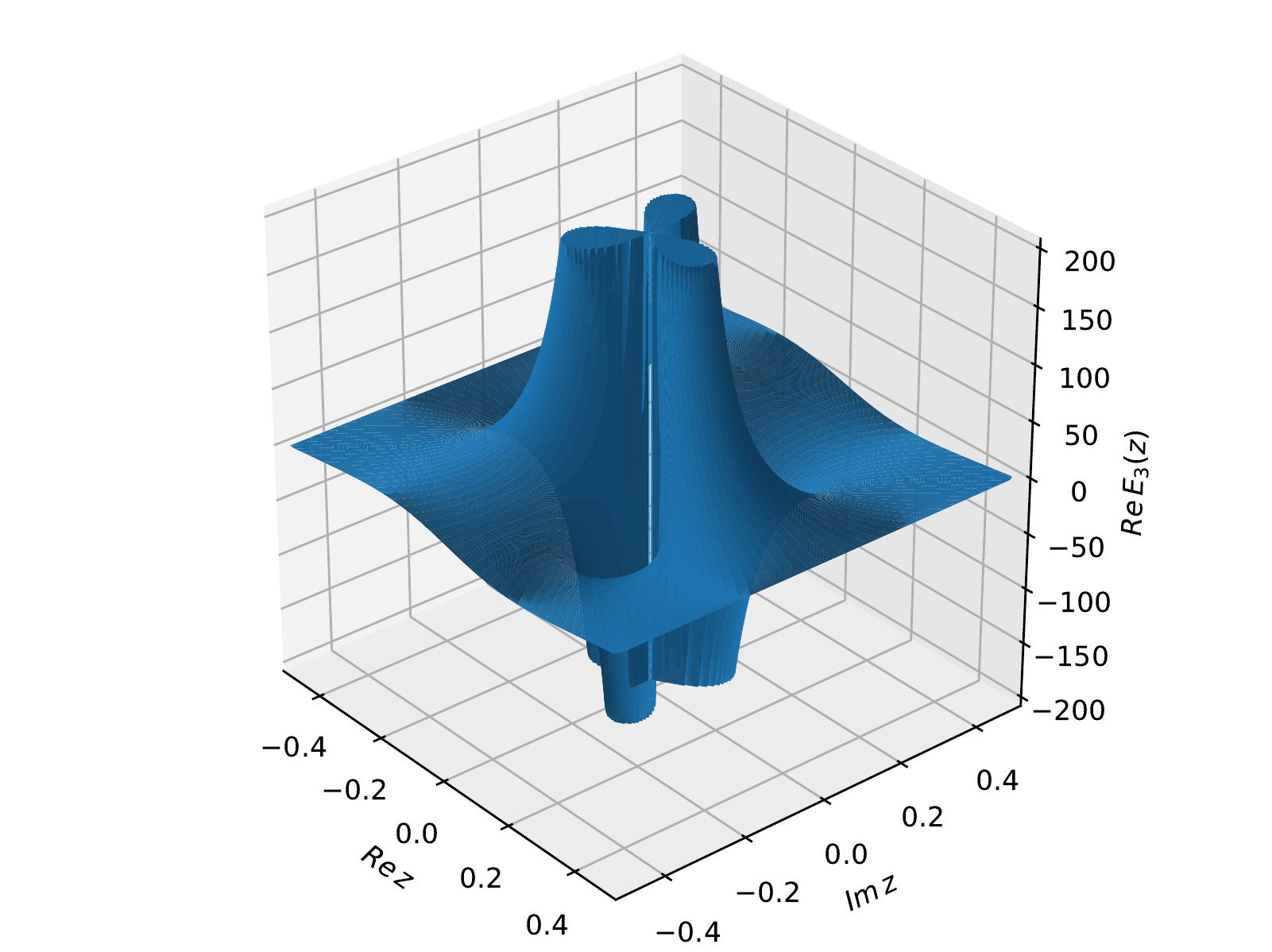}
\includegraphics[clip, trim=25mm 0mm 15mm 5mm, width=.44\textwidth]{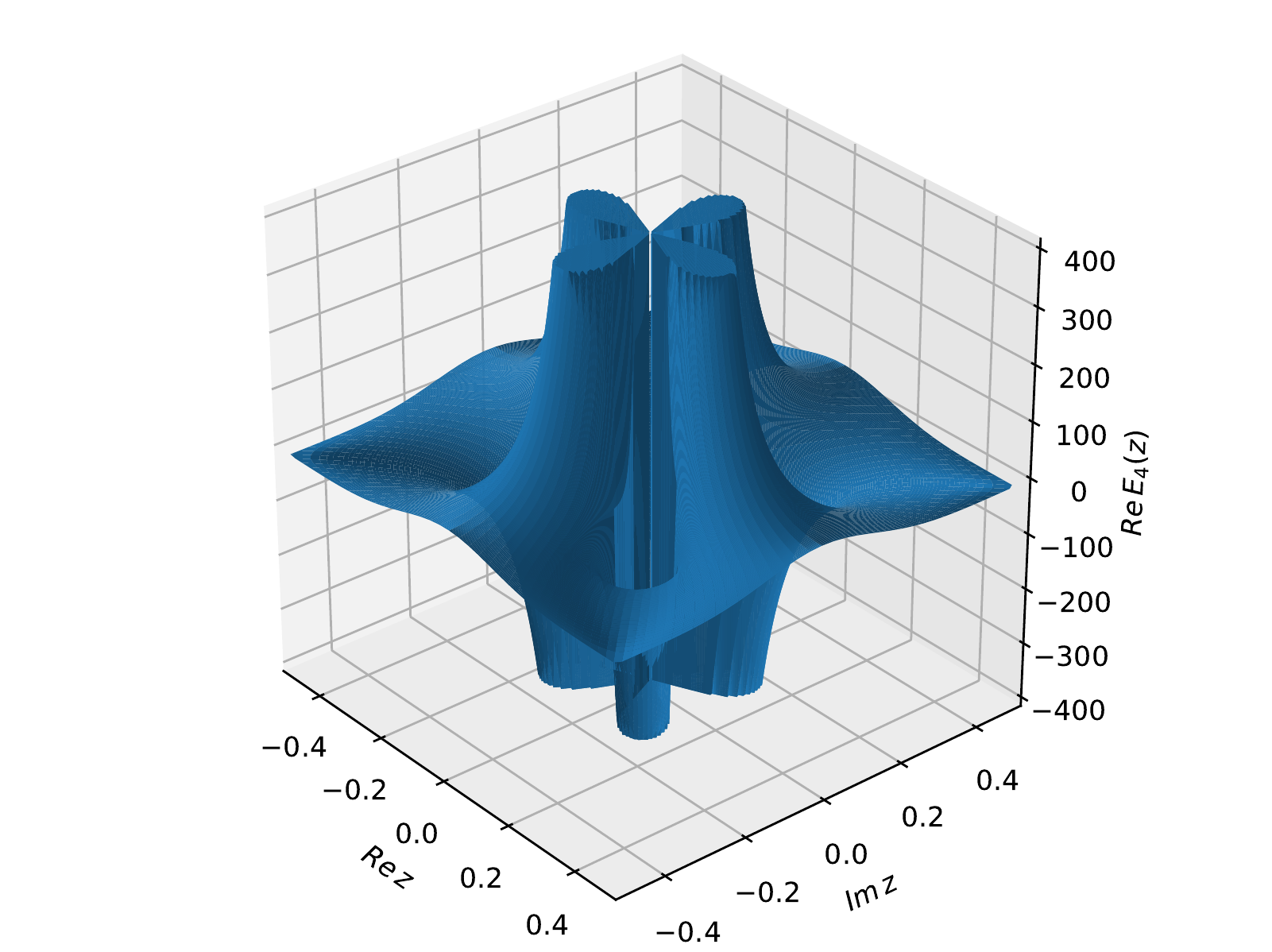} 
\caption{Plots of Eisenstein function $E_n$ for $n=3, 4$ for the square lattice, where $\omega_1=1$ and $\omega_2=i$.} 
\label{fig:eis_plots}
\end{figure}

It is difficult to explicitly characterize the geometric meaning of the structural sums. The elements $e_{\mathbf m}$ can be treated as weighted moments (integrals
over the cell) of the correlation functions \cite{GluMitNaw2017Book}. Use of $e_{\mathbf m}$ allows to avoid huge computations of the correlation functions and compute implicitly their weighted moments of high orders. 
 One can also consider structural sums as {\it generalized directed distances}. Indeed, each term of the sum~\eqref{eq:eSum} can be intuitively interpreted as a chain of reciprocals of powers of directed distances (complex numbers). Hence, a particular sum is some kind of a summary characteristic of the network formed by disks.
The higher the degree of function $E_n$ is, the more complex form it has (see Fig.~\ref{fig:eis_plots}).


\subsection{Algorithms for structural sums}
\label{sect:computatons}
There have been significant updates of algorithms for structural sums in recent years. For example, the paper~\cite{NawComp2017} in detail describes algorithms and methods for efficient computations of {\it discrete multidimensional convolutions of functions} with applications to structural sums, being the special case of such multiple convolution. The application of the specific vector-matrix representation of a convolution, presented therein, significantly reduces computational complexity of calculations and allows reusing intermediate results cached during the computations of a large number of sums. Such an approach can be directly applied to the structural sums feature vector.
On the other hand, the relations~\eqref{eq:2.18} are the direct consequence of  theorems and algorithms reported in~\cite{NawSymb2016}.
The very same paper introduces methods for reducing dependent sums arising in the set~$G$.
%
%

\section{Feature vector construction}
\label{sect:feature_vector}

One can consider the set $G$ as the base of the general form of the feature vector of random composites modelled by non-overlapping disks on the plane. The set $\mathcal M_e$ discussed in section~\ref{sect:theory} is infinite, therefore applications require finite approximations. Let $M_q$ be the set of all structural sums of order $q$. By the
virtue of relations~\eqref{eq:2.18}, consecutive sets $M_q$ can be generated
iteratively as follows:
\begin{align*}
M_1 =& \{e_{2}\}, \quad M_2 = \{e_{2, 2}\}, \quad M_3 = \{e_{2, 2, 2}, e_{3, 3}\}\\
M_4 =& \{e_{2, 2, 2, 2}, e_{2, 3, 3}, e_{3, 3, 2}, e_{4, 4} \}\\
M_5 =& \{e_{2, 2, 2, 2, 2}, e_{2, 2, 3, 3}, e_{2, 3, 3, 2}, e_{2, 4, 4}, e_{3, 3, 2, 2}, e_{3, 4, 3},\\
    &  e_{4, 4, 2}, e_{5, 5} \}.\\
M_6 =& \{e_{2, 2, 2, 2, 2, 2}, e_{2, 2, 2, 3, 3}, e_{2, 2, 3, 3, 2}, e_{2, 2, 4, 4}, e_{2, 3, 3, 2, 2},\\
    &  e_{2, 3, 4, 3}, e_{2, 4, 4, 2}, e_{2, 5, 5}, e_{3, 3, 2, 2, 2}, e_{3, 3, 3, 3}, e_{3, 4, 3, 2},\\
    &   e_{3, 5, 4}, e_{4, 4, 2, 2}, e_{4, 5, 3}, e_{5, 5, 2}, e_{6, 6} \}.
\end{align*}
Note that $G = \bigcup_{j=1}^\infty M_j$.
Moreover, $G$ involves dependent sums. Indeed, the following relation~\cite{NawSymb2016}:
\begin{equation}
  \label{eq:mirr1}
  \begin{array}{lll}
e^{\nu_0,\nu_1,\nu_2,\ldots,\nu_n}_{p_1,p_2,p_3,\ldots,p_n}=(-1)^\alpha\mathbf{C}^{n+1}e^{\nu_n,\nu_{n-1},\nu_{n-2},\ldots,\nu_1}_{p_n,p_{n-1},p_{q-2},\ldots,p_1},
\end{array}
\end{equation}
where $\alpha=\sum^n_{j=1}p_j$, describes a relationship between so-called {\it mirror structural sums}. One can apply this relation in order to consider only independent sums. As it turns out, \eqref{eq:mirr1} reduces elements of $M_q$ almost by half.
Let $G_q$ denote subset of $M_q$, containing only independent
sums. For instance, the set~$G_6$
takes the following form:
\begin{align*}
G_6 =& \{e_{2, 2, 2, 2, 2, 2}, e_{2, 3, 3, 2, 2},e_{2, 4, 4, 2}, e_{3, 3, 2, 2, 2}, e_{3, 3, 3, 3}, \\
& e_{3, 4, 3, 2}, e_{4, 4, 2, 2}, e_{4, 5, 3}, e_{5, 5, 2}, e_{6, 6} \}.
\end{align*}
Hence, we define {\it structural sums feature vector of order q} as follows:
\begin{equation}
  \label{eq:features}
	X_q = \{G_1, G_2, G_3, \ldots, G_q \}.
\end{equation}
For example, the feature vector of order 4 has the following form:
$$X_4 = \{e_{2}, e_{2, 2}, e_{2, 2, 2}, e_{3, 3}, e_{2, 2, 2, 2}, e_{2, 3, 3}, e_{4, 4} \}.$$


Since the elements of $X_q$ are complex numbers, one can use different vectors derived from $X_q$, for instance:
\begin{itemize}
 \setlength\itemsep{0em}
\item $|X_q|$ -- absolute values of elements of $X_q$;
\item ${\mathcal Re}X_q$ -- real parts of elements of $X_q$;
\item ${\mathcal Im}X_q$ -- imaginary parts of elements of $X_q$;
\item ${\mathcal Arg}X_q$ -- arguments (angles) of elements of $X_q$;
\item $\left[{\mathcal Re}X_q, {\mathcal Im}X_q\right]$  -- both real and imaginary parts of elements of $X_q$.
\end{itemize}

Note that we described the construction of the general form of the feature vector. In applications, one can select a certain subset of structural sums that gives satisfactory results. For instance, the following set:
$$X'_q = \{e_{p, p} \; : \; 2\leq p\leq q \}$$
is the smallest subset involving Eisenstein functions $E_n$ ($n=2, 3,\ldots,q$). 
Some particular datasets may reveal another corelations between sums, allowing further elimination of dependent parameters. For example, considering models of macroscopically isotropic composites, one can apply so-called Keller's identity (for more details, see~\cite{MitRyl2012}).



\section{Classification experiments}
\label{sect:examples}
%

In order to explore structural sums as an information carrier, in present section we discuss two examples of classification of simulated composites based on the general structural sums vector's approximations. These examples serve to illustrate two important issues. Firstly, we want to answer the question, whether the higher-order sums are worth analysing or all the information is encoded in the lower-order parameters. Secondly, it will be shown that different kind of data may require the use of a specific modification of the vector $X_q$. The examples discuss composites with inclusions modelled by distributions of disks and distributions of shapes formed by disks (clusters). 

All of the examples follow the same framework, known in machine learning as {\it classification problem}, consisting in taking input vectors and deciding which of $c$ classes
they belong to, based on {\it training} from exemplars of each class~\cite{Marsland2014}. Let $C_j$ ($j=1,2, 3,\ldots,c$) denote considered classes of distributions. For each $C_j$ we generate a set of 100 samples drawn from a given distribution. In order to create a {\it training set}, we randomly pick $k$ samples from each class. The training set is used to build a data model. In order to decide how well the algorithm has learnt using a given approximation of feature vector, we predict the outputs on a {\it test set} consisting of $(100-k)c$ remaining samples (i.e. the portion of data not seen in the training set). Note that in our examples we use $k\leq 40$, hence we are dealing with a relatively large test set ($60\%$-$96\%$ of the original dataset).
For the feature vector, we use different modifications of the structural sums vector~\eqref{eq:features} of varying order $q$ ($q=1,2,\ldots,10$).

\begin{figure}[!t]
\centering
\includegraphics[clip, trim=0mm 0mm 0mm 0mm, width=.6\textwidth]{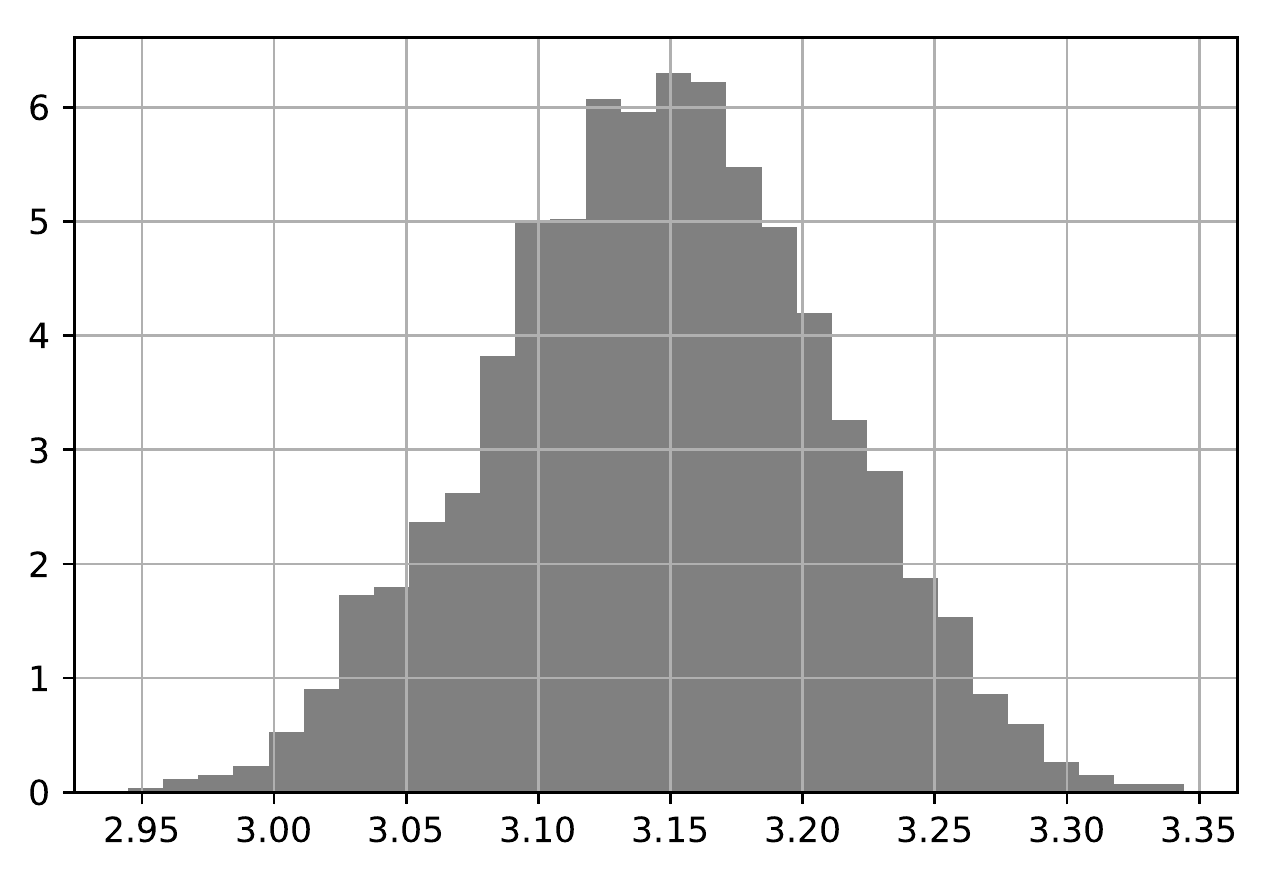} 
\caption{\footnotesize{Histogram of values of sum $e_2$ computed for 2000 samples of $N=100$ disks of concentration $\nu=0.45$ generated via random walk in the square cell.
}}
\label{fig:e2hist}
\end{figure}

The data model we use is based on a simple Gaussian Naive Bayes classification algorithm, being capable of handling multiple classes directly,
implemented in Scikit-Learn Python module~\cite{scikitlearn}. The Naive Bayes  classifier also assumes that features are independent of each other. In general, this assumption may not be true, however yields computational simplification. Moreover, the likelihood of the features is assumed to be Gaussian among a given class, which seems to be justified by an observation that the values of structural sums tend to be normally distributed (see Fig.~\ref{fig:e2hist}).
For more details on the Naive Bayes classifier, see the Scikit-Learn module documentation or \cite{Marsland2014}.

Consider the accuracy of the prediction as the ratio of correctly classified samples to all the samples. In each case we build 10 random splits of the set, as described above, and the final score is the arithmetic mean of resulting accuracy values. Our goal is to observe how the accuracy changes depending on the type of modification of the feature vector, the order $q$, and the size $kc$ of a training set.

\subsection{Circular inclusions}
\label{sect:disks}

In the first experiment, we generate various distributions of disks on the plane and verify whether they are distinguishable by the structural sums feature vector. In order to generate sample data, we use a specific random walk algorithm with varying parameters. 

\begin{figure}[ht]
\centering
\includegraphics[clip, trim=0mm 0mm 0mm 0mm, width=.6\textwidth]{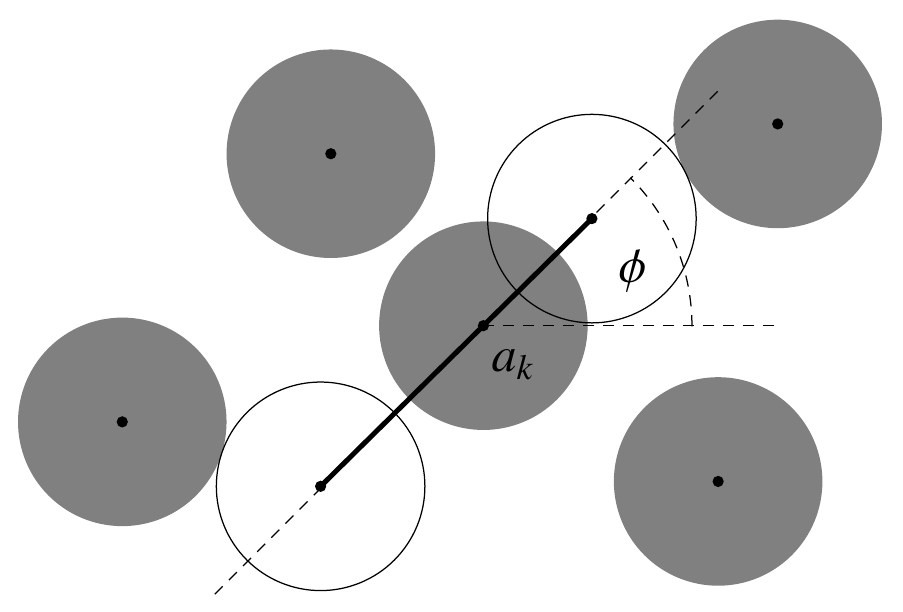} 
\caption{\footnotesize{Illustration of the random move of disk $a_k$ in a direction $\phi$ on the distance $d$ in the direction $\phi$ up to  $d_{max}$ or in the direction $(\phi-\pi)$ up to $d_{min}$.
}}
\label{fig:rand_walk_1}
\end{figure}

Let us describe the general Markov-chain (MC) protocol for Monte Carlo simulations.
Assume that initially all the centres are placed in the cell, for instance, as a regular array of disks or as a result of the Random Sequential Adsorption (RSA) protocol, where consecutive objects are placed randomly in the cell, rejecting those that overlap previously absorbed ones. For each $a_k$ ($k = 1,2,\ldots,N$) we execute the following procedure. First, for a given centre $a_k$, a step direction $\phi$ is chosen as a realization of the random variable uniformly distributed in the interval $(0,\pi)$. Furthermore, the values $d_{\max}>0$ and $d_{\min}<0$ are computed in such a way that the disk can move along the directions, respectively, $\phi$ or $\phi+\pi$ without collisions with other disks (see Fig.~\ref{fig:rand_walk_1}). If collisions do not occur, $a_k$ can move up to the boundary of the cell $Q$.
Otherwise, the point $a_k$ moves along the direction $\phi$ (or $\phi+\pi$) with the distance $d$ considered to be a realization of a random variable distributed in $(d_{\min},d_{\max})$, taking $d = d_{min} + (d_{max}-d_{min})Z$. Here, $Z$ is a random variable distributed in the interval $[0, 1]$. Negative values of $d$ correspond to the direction $\phi+\pi$.
Hence, every centre obtains a new complex coordinate $a'_k$ after the performed step. 

This move is repeated with new coordinates for each $k = 1,2,\ldots,N$. 
We say that a {\it cycle} is performed if $k$ runs from $1$ to $N$. 
The number of cycles, denoted by $t$, corresponds to time of random walks of disks. A location of points $a_k$ at time $t$ forms a probabilistic event described by distribution $\mathcal U_{t}(\nu)$.  After a sufficient number of walks $t$, the obtained location of the centres can be considered a statistical realization of the distribution $\mathcal U_{\infty}(\nu)$.

\begin{figure}[!t]
\centering
\begin{tabular}{cccc}


&
{\footnotesize $Z_1$}\includegraphics[clip, trim=14mm 10mm 5mm 0mm, width=.1\textwidth]{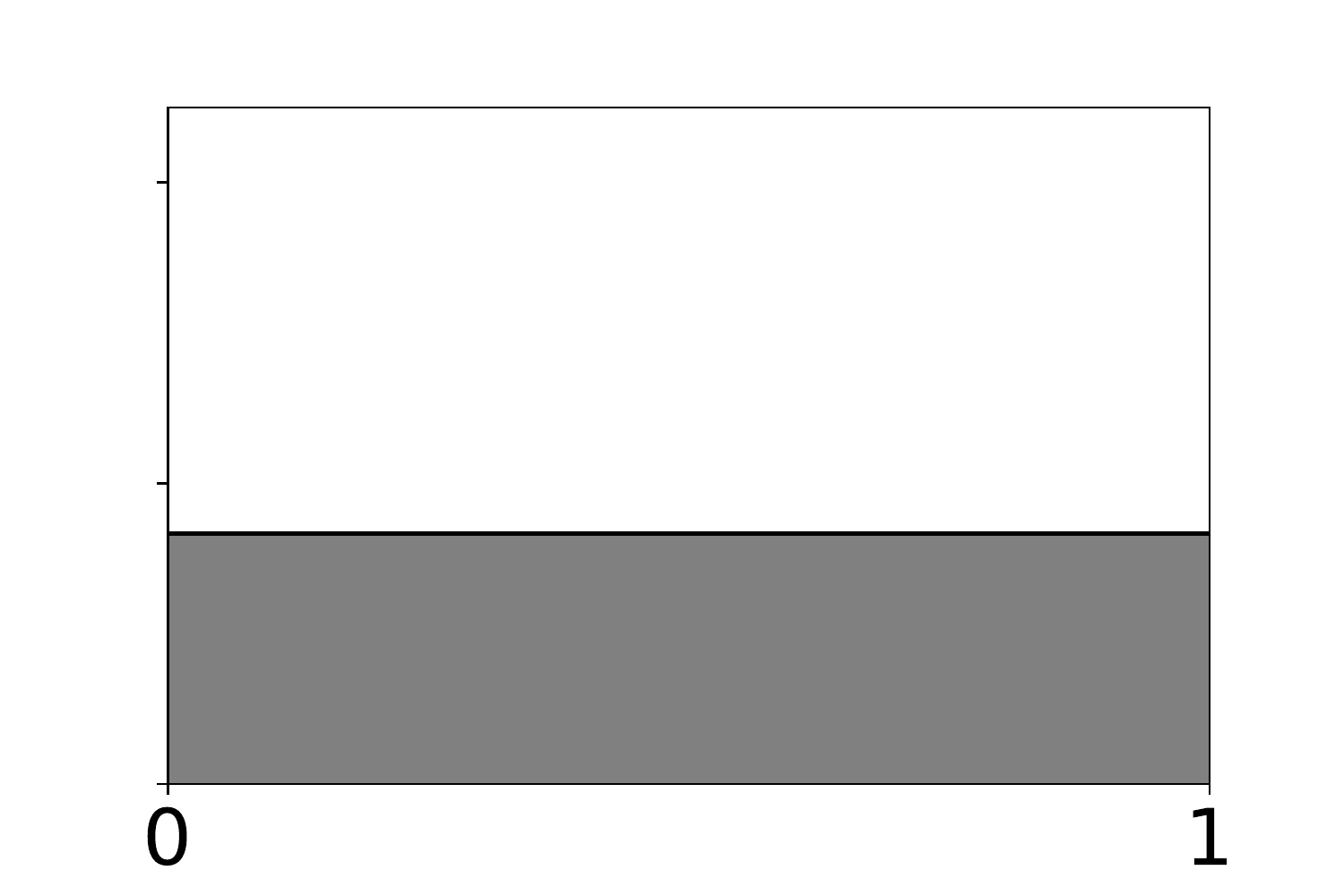} &
{\footnotesize $Z_2$}\includegraphics[clip, trim=14mm 10mm 5mm 0mm, width=.1\textwidth] {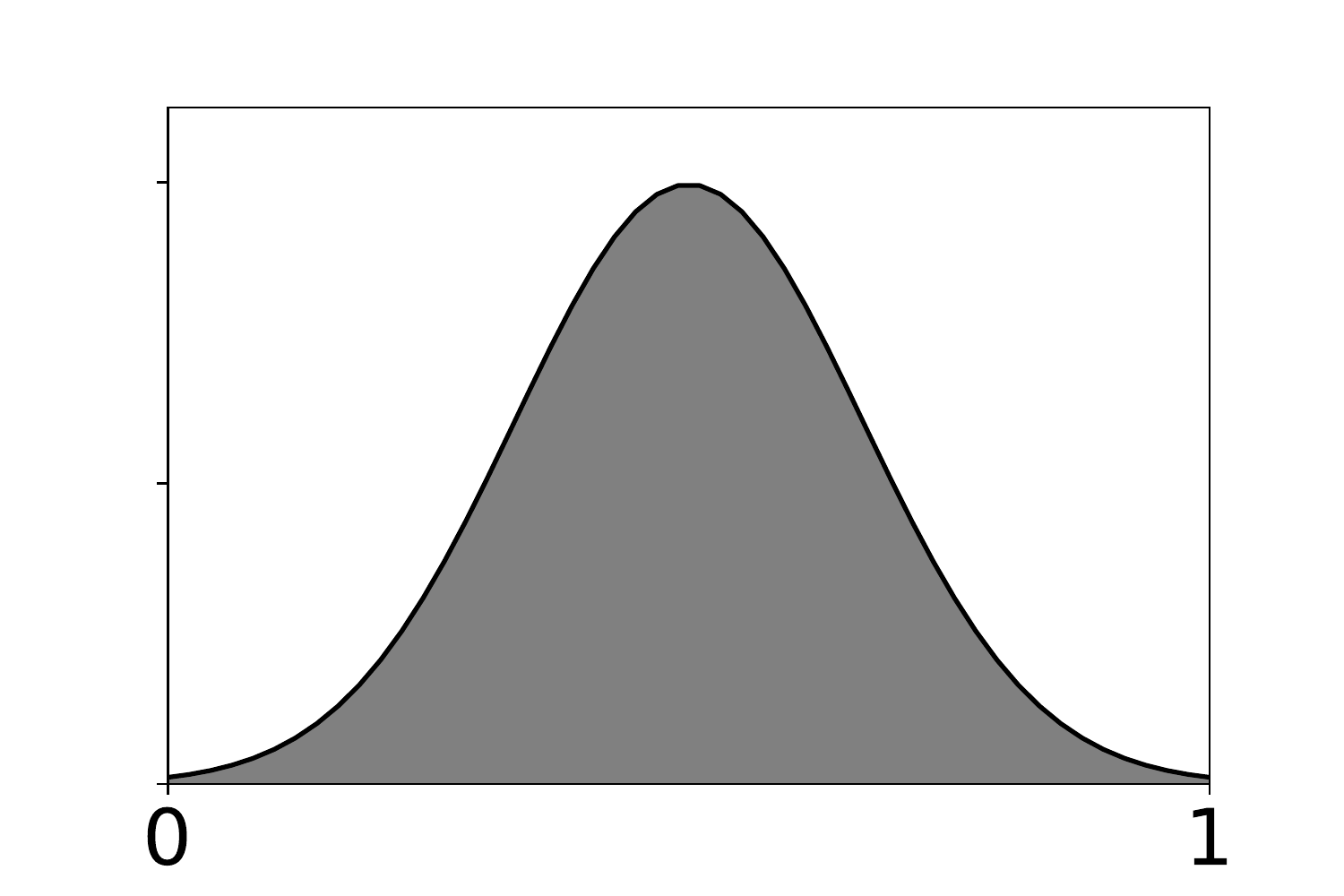} &
{\footnotesize $Z_3$}\includegraphics[clip, trim=14mm 10mm 5mm 0mm, width=.1\textwidth]{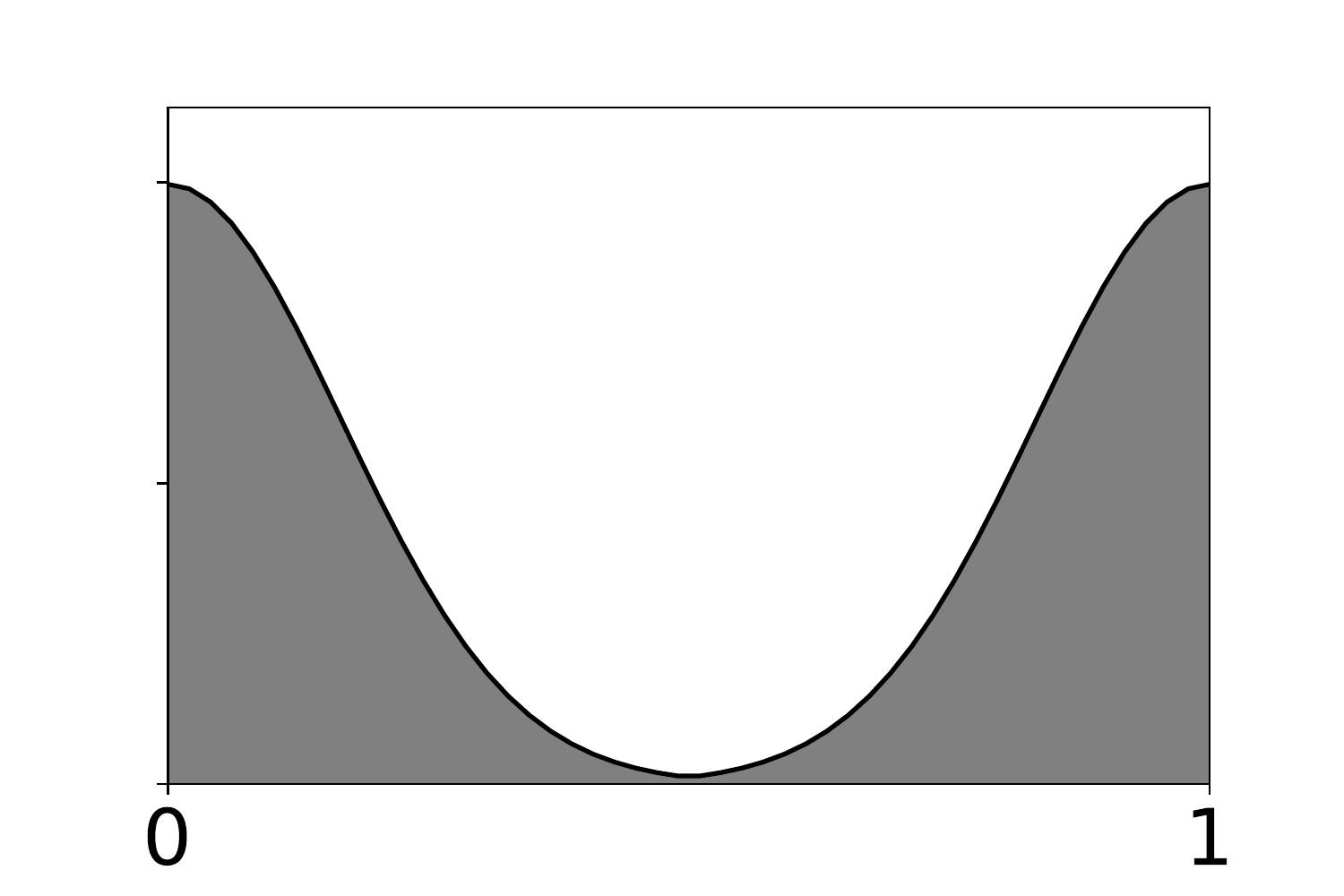} \\

{\rotatebox[origin=l]{90}{{\footnotesize \hspace{0.65cm} identical radii}}} &
\includegraphics[clip, trim=14mm 10mm 5mm 0mm, width=.28\textwidth]{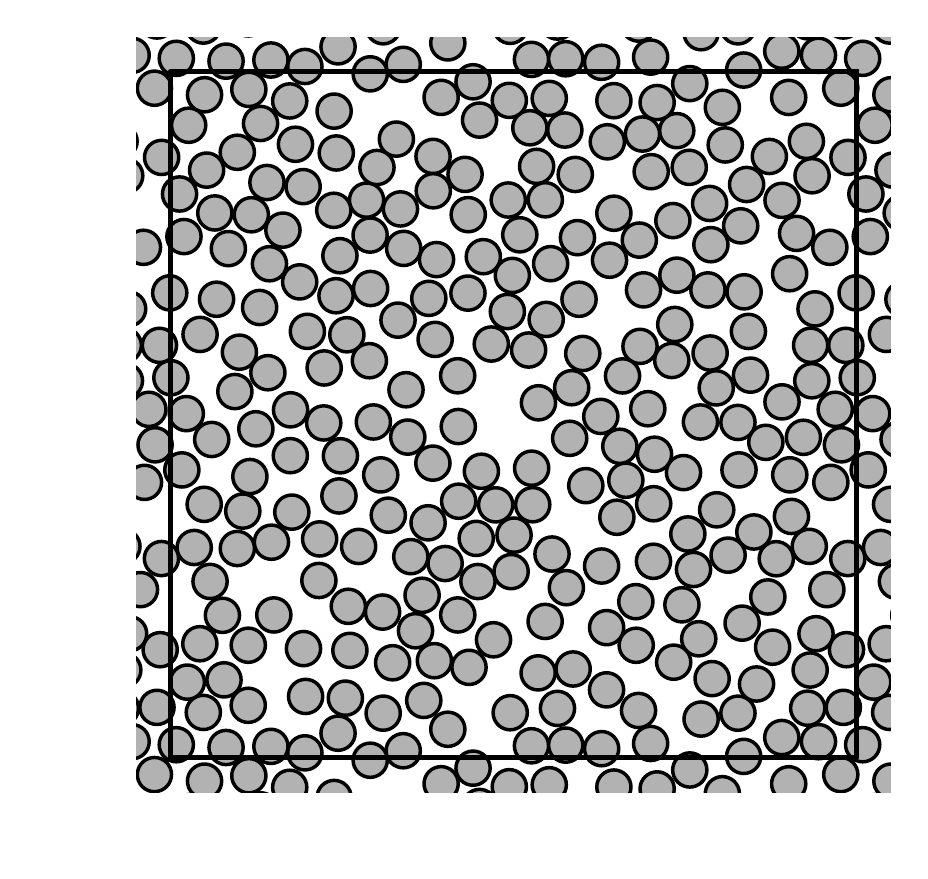} &
\includegraphics[clip, trim=14mm 10mm 5mm 0mm, width=.28\textwidth]{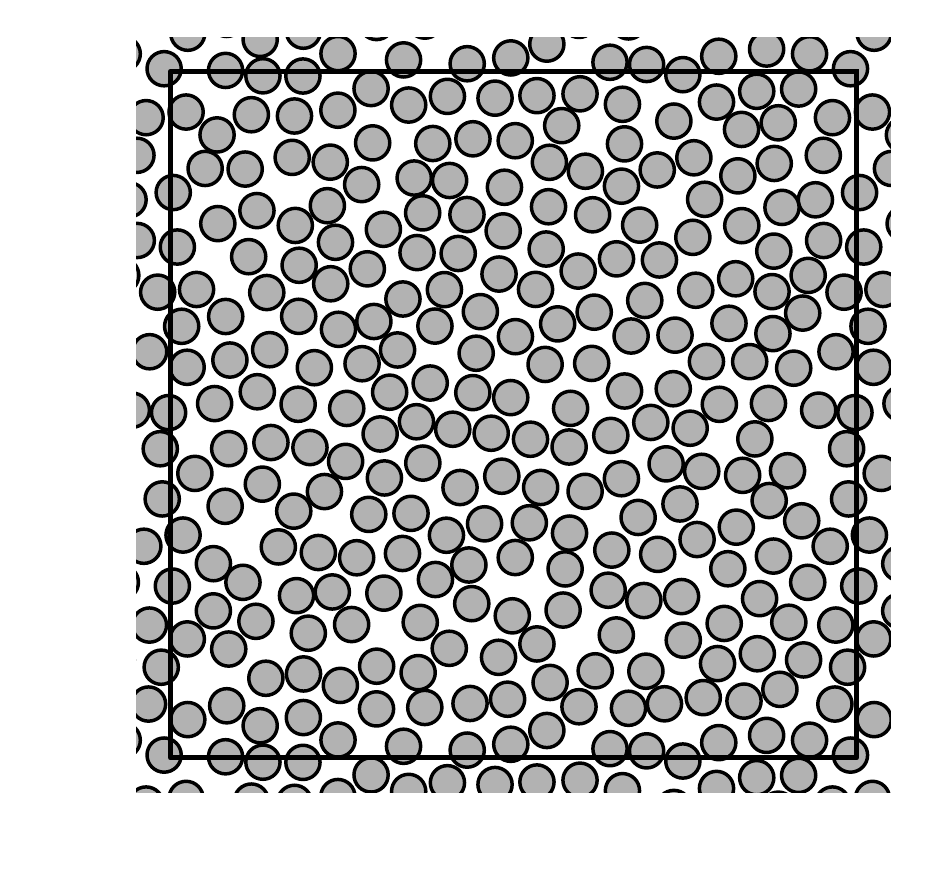} &
\includegraphics[clip, trim=14mm 10mm 5mm 0mm, width=.28\textwidth]{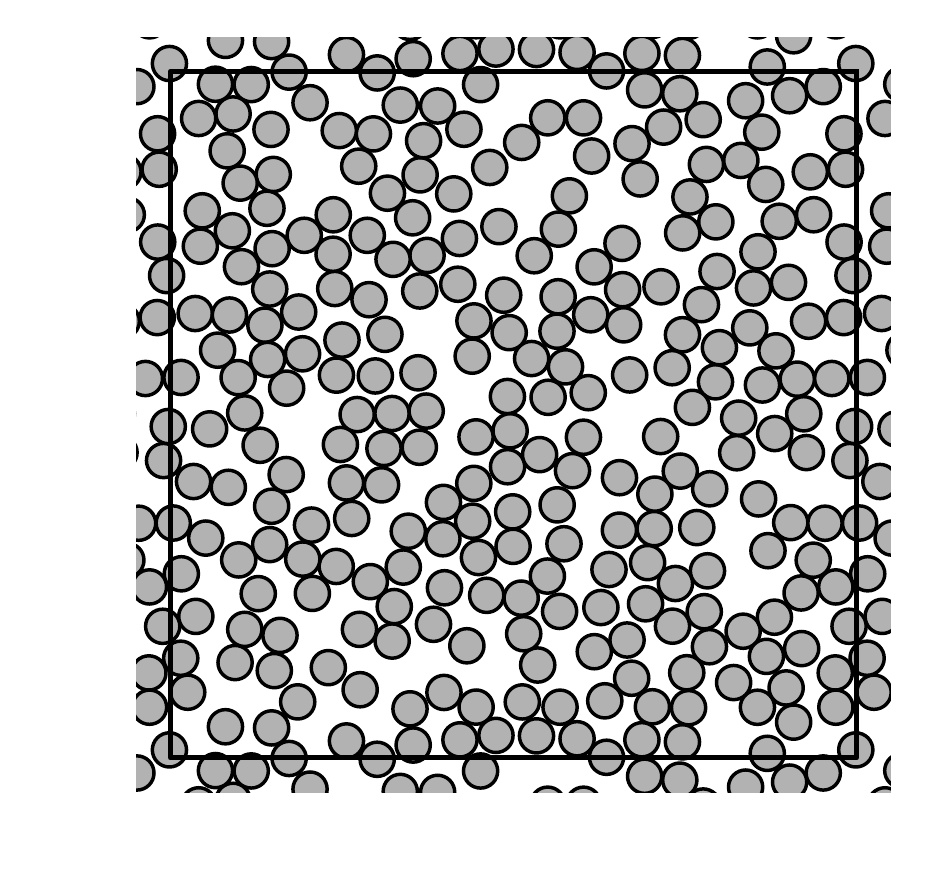} \\

{\rotatebox[origin=l]{90}{{\footnotesize \hspace{-0.15cm} uniformly distributed radii}}} &
\includegraphics[clip, trim=14mm 10mm 5mm 0mm, width=.28\textwidth]{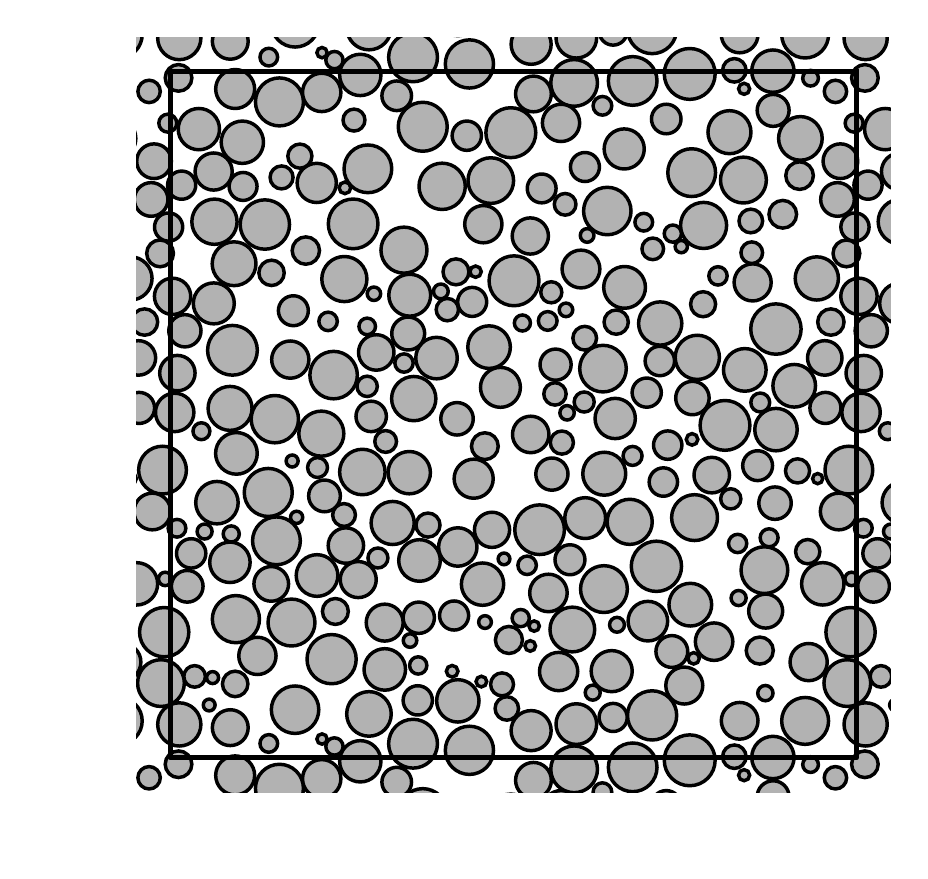} &
\includegraphics[clip, trim=14mm 10mm 5mm 0mm, width=.28\textwidth]{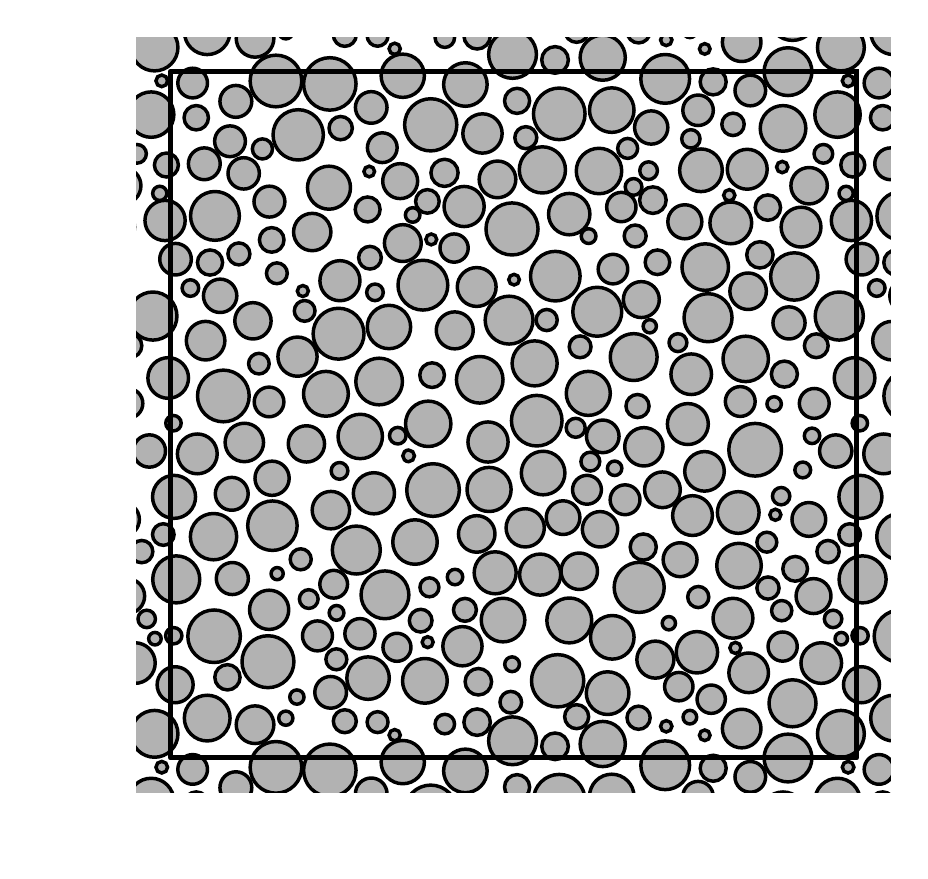} &
\includegraphics[clip, trim=14mm 10mm 5mm 0mm, width=.28\textwidth]{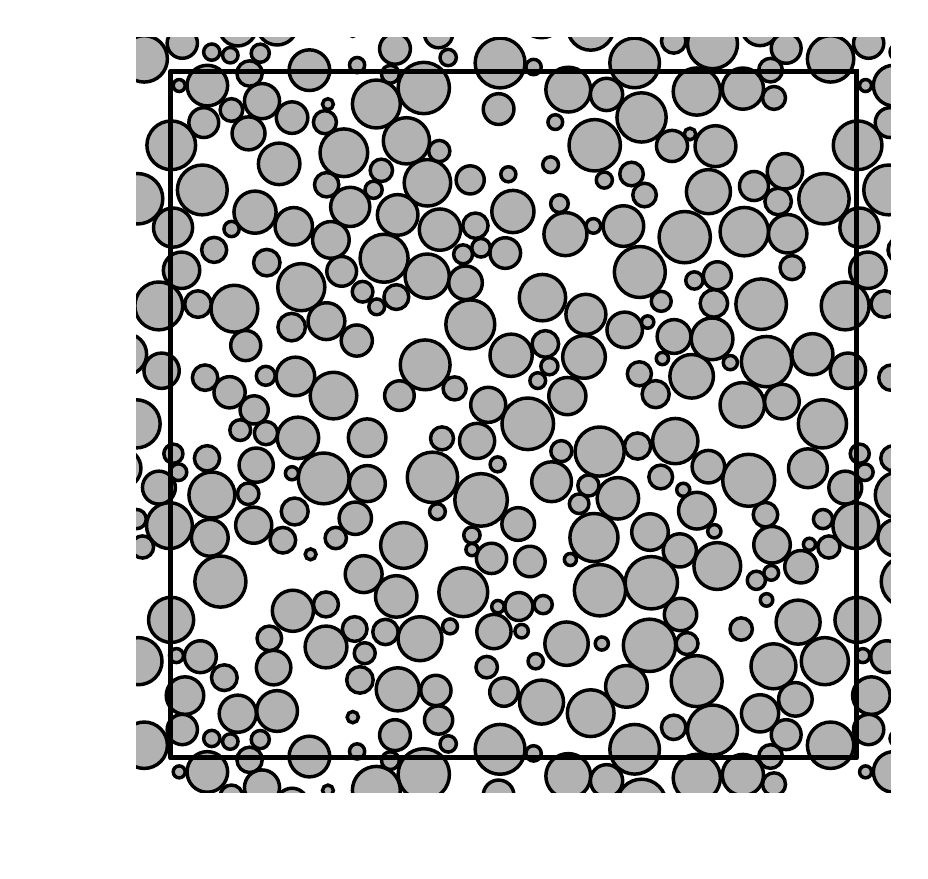} \\

{\rotatebox[origin=l]{90}{{\footnotesize \hspace{-0.1cm} normally distributed radii}}} &
\includegraphics[clip, trim=14mm 10mm 5mm 0mm, width=.28\textwidth]{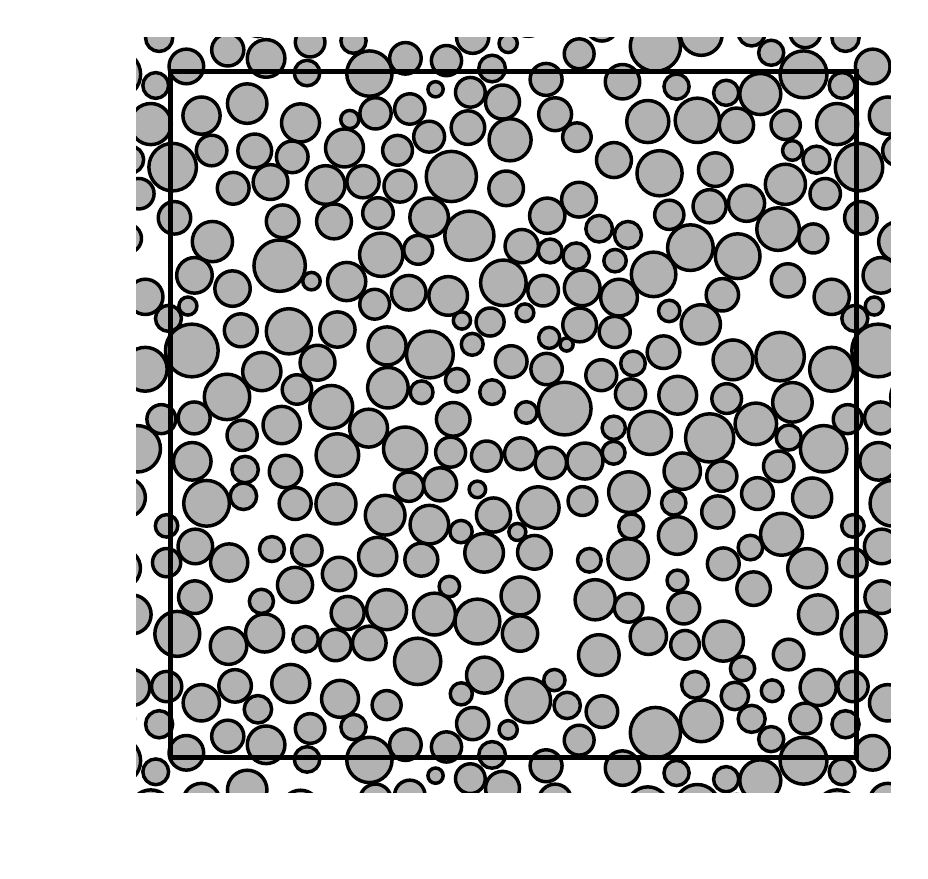} &
\includegraphics[clip, trim=14mm 10mm 5mm 0mm, width=.28\textwidth]{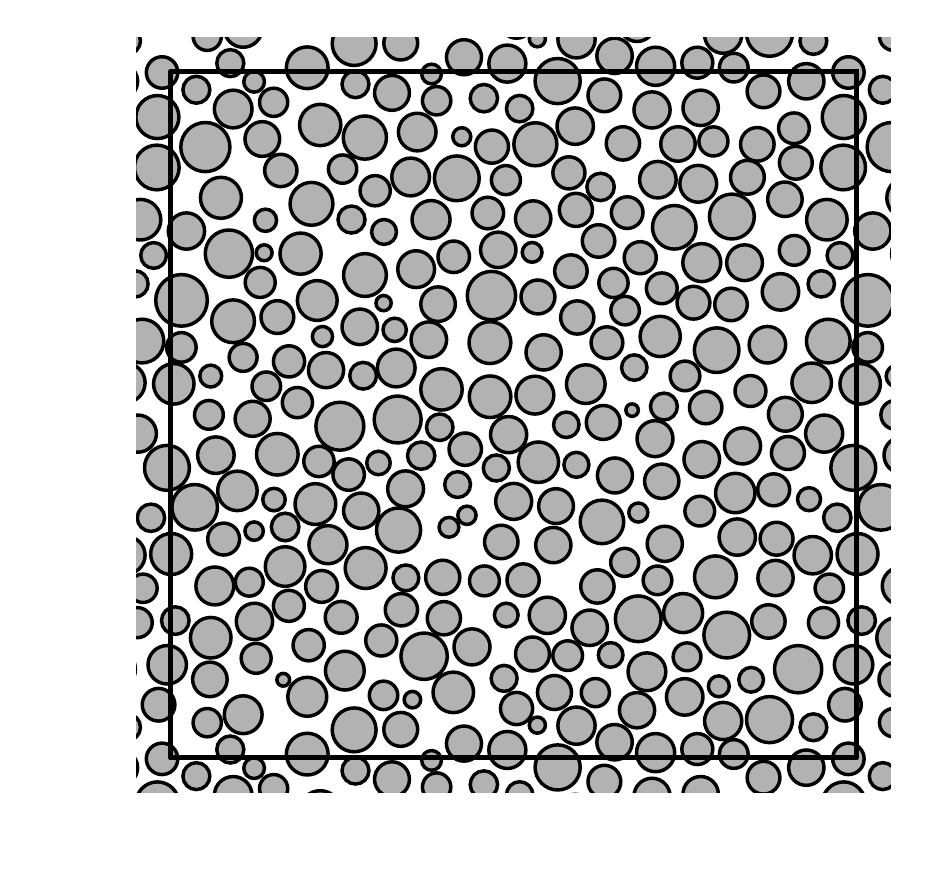} &
\includegraphics[clip, trim=14mm 10mm 5mm 0mm, width=.28\textwidth]{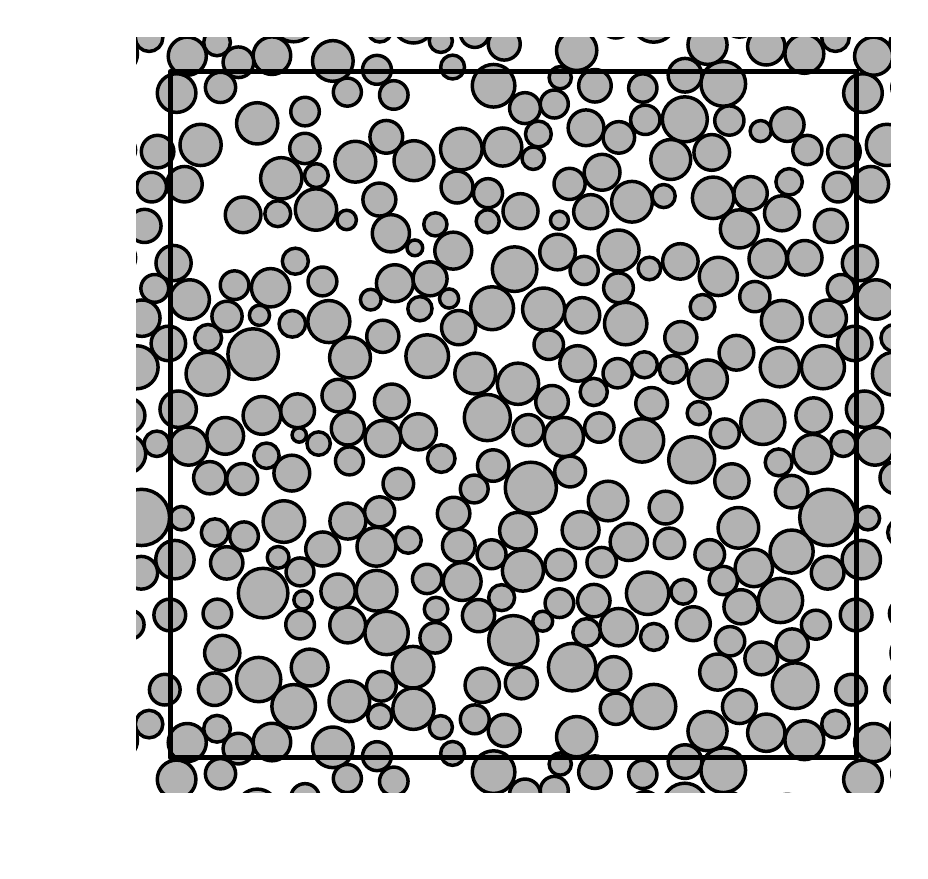} \\

\end{tabular}
\caption{Instances of distributions of disks used in classification problems.}
\label{fig:distributions}
\end{figure}

One can obtain various classes of distributions of disks on the plane via altering two parameters: random variable $Z$ and the distribution of radii of disks. 
In our experiments we consider following three examples in place of variable~$Z$.
Let $Z_1$ be the random variable uniformly distributed in the interval $[0, 1]$ and let
$$Z_2 = \frac{Z_{\mathcal N}}{6}+\frac12$$
$$Z_3 = frac\left(\frac{Z_{\mathcal N}}{6}+1\right),$$
 where $Z_{\mathcal N}$ is the random variable with the standard normal distribution $\mathcal N(\mu=0,\sigma^2=1)$ truncated to the range $[-3, 3]$ and $frac(x)$ is the fractional part of $x$. The PDFs of considered random variables are shown at the top of Fig.~\ref{fig:distributions}. 
%
%
%
Together with $Z_j$ ($j=1, 2, 3$), we look into different distributions of radii of disks (i.e. identical disks, uniformly distributed radii and normally distributed radii). Hence, we end up with 9 classes being combinations of those two parameters. Our simulations are based on distributions of 256 disks of concentration $\nu=0.5$ generated by the MC protocol described above (see examples in Fig.~\ref{fig:distributions}).

The results of classification, in varying both the size of training data and the order $q$ of the modified feature vector~$X_q$,   are presented in Table~\ref{tab:acc_distr}. One can see that very good results are obtained for a relatively small training set (7\%-19\% of the dataset) for vectors $|X_q|$ and ${\mathcal Re}X_q$ of orders 8-10. Note that a purely random classifier has the accuracy approximately equal 0.11.

\begin{table}[!t]
\caption{Accuracy of classification of distributions of disks for $|X_q|$, ${\mathcal Re}X_q$, ${\mathcal Im}X_q$, and ${\mathcal Arg}X_q$.}
\label{tab:acc_distr}
\setlength{\tabcolsep}{3pt}
\centering\scriptsize
\begin{tabular}{cccccccccccc}
\cline{1-12}
\multicolumn{2}{c}{train}  & \multicolumn{10}{c}{order of feature vector} \\ \cline{3-12}

\multicolumn{2}{c}{size} &        1  &        2  &        3  &        4  &        5  &        6  &        7  &        8  &        9  &        10 \\   \hline 

\multicolumn{12}{l}{$|X_q|$}\\
& 4\%         &  0.120 &  0.349 &  0.427 &  0.500 &  0.537 &  0.556 &  0.690 &  0.773 &  0.840 &  0.858 \\
& 7\%         &  0.129 &  0.340 &  0.427 &  0.544 &  0.593 &  0.637 &  0.772 &  0.861 &  0.902 &  0.918 \\
&10\%         &  0.135 &  0.350 &  0.457 &  0.559 &  0.618 &  0.663 &  0.800 &  0.883 &  0.927 &  0.941 \\
&13\%         &  0.135 &  0.343 &  0.463 &  0.570 &  0.626 &  0.670 &  0.806 &  0.895 &  0.925 &  0.936 \\
&16\%         &  0.140 &  0.348 &  0.472 &  0.585 &  0.650 &  0.700 &  0.841 &  0.913 &  0.944 &  0.954 \\
&19\%         &  0.141 &  0.354 &  0.476 &  0.600 &  0.658 &  0.709 &  0.843 &  0.915 &  0.948 &  0.957 \\
&22\%         &  0.142 &  0.354 &  0.479 &  0.599 &  0.666 &  0.718 &  0.850 &  0.924 &  0.953 &  0.961 \\
&25\%         &  0.140 &  0.338 &  0.483 &  0.612 &  0.682 &  0.734 &  0.864 &  0.935 &  0.958 &  0.965 \\
&28\%         &  0.142 &  0.345 &  0.489 &  0.610 &  0.676 &  0.727 &  0.864 &  0.934 &  0.958 &  0.964 \\
&31\%         &  0.147 &  0.339 &  0.492 &  0.609 &  0.679 &  0.731 &  0.862 &  0.934 &  0.960 &  0.965 \\
&34\%         &  0.136 &  0.344 &  0.490 &  0.614 &  0.676 &  0.730 &  0.867 &  0.932 &  0.959 &  0.963 \\
&37\%         &  0.140 &  0.338 &  0.494 &  0.606 &  0.675 &  0.726 &  0.864 &  0.932 &  0.956 &  0.961 \\
&40\%         &  0.138 &  0.334 &  0.491 &  0.611 &  0.686 &  0.740 &  0.878 &  0.937 &  0.961 &  0.966 \\

\multicolumn{12}{l}{${\mathcal Re}X_q$}\\
& 4\%         &  0.120 &  0.350 &  0.428 &  0.501 &  0.538 &  0.561 &  0.689 &  0.777 &  0.833 &  0.851 \\
& 7\%         &  0.129 &  0.340 &  0.427 &  0.544 &  0.591 &  0.632 &  0.772 &  0.858 &  0.898 &  0.914 \\
&10\%         &  0.136 &  0.351 &  0.457 &  0.560 &  0.617 &  0.664 &  0.802 &  0.881 &  0.925 &  0.941 \\
&13\%         &  0.135 &  0.343 &  0.463 &  0.570 &  0.625 &  0.670 &  0.814 &  0.896 &  0.926 &  0.939 \\
&16\%         &  0.140 &  0.348 &  0.472 &  0.584 &  0.644 &  0.695 &  0.838 &  0.913 &  0.941 &  0.954 \\
&19\%         &  0.141 &  0.356 &  0.476 &  0.600 &  0.662 &  0.711 &  0.846 &  0.917 &  0.949 &  0.960 \\
&22\%         &  0.141 &  0.354 &  0.479 &  0.599 &  0.667 &  0.715 &  0.853 &  0.922 &  0.953 &  0.962 \\
&25\%         &  0.140 &  0.338 &  0.483 &  0.612 &  0.680 &  0.729 &  0.863 &  0.936 &  0.959 &  0.967 \\
&28\%         &  0.142 &  0.345 &  0.489 &  0.610 &  0.679 &  0.729 &  0.865 &  0.936 &  0.959 &  0.964 \\
&31\%         &  0.147 &  0.339 &  0.492 &  0.609 &  0.679 &  0.733 &  0.863 &  0.933 &  0.959 &  0.964 \\
&34\%         &  0.137 &  0.345 &  0.490 &  0.614 &  0.673 &  0.729 &  0.869 &  0.934 &  0.958 &  0.962 \\
&37\%         &  0.138 &  0.338 &  0.494 &  0.606 &  0.673 &  0.728 &  0.862 &  0.930 &  0.956 &  0.961 \\
&40\%         &  0.138 &  0.335 &  0.490 &  0.610 &  0.687 &  0.738 &  0.876 &  0.938 &  0.960 &  0.966 \\

\multicolumn{12}{l}{${\mathcal Im}X_q$}\\
&16\%         &  0.139 &  0.139 &  0.147 &  0.181 &  0.223 &  0.256 &  0.312 &  0.378 &  0.470 &  0.554 \\
&28\%         &  0.146 &  0.146 &  0.151 &  0.184 &  0.233 &  0.271 &  0.341 &  0.426 &  0.542 &  0.637 \\
&40\%         &  0.140 &  0.140 &  0.151 &  0.188 &  0.241 &  0.274 &  0.346 &  0.441 &  0.551 &  0.654 \\

\multicolumn{12}{l}{${\mathcal Arg}X_q$}\\
&16\%         &  0.139 &  0.139 &  0.141 &  0.138 &  0.144 &  0.143 &  0.158 &  0.186 &  0.219 &  0.277 \\
&28\%         &  0.145 &  0.145 &  0.144 &  0.141 &  0.151 &  0.152 &  0.169 &  0.202 &  0.250 &  0.310 \\
&40\%         &  0.141 &  0.141 &  0.135 &  0.138 &  0.145 &  0.152 &  0.172 &  0.214 &  0.256 &  0.336 \\
\end{tabular}
\end{table}
Both a high accuracy of $|X_q|$ and a poor performance of ${\mathcal Arg}X_q$ seem to be justified. Differences between presented classes of distributions appear to be strongly related to distances between inclusions. On the other hand, since the samples show the uniformity in all orientations (i.e. isotropy), the vector ${\mathcal Arg}X_q$ seems to be less relevant. It was also observed that in case of such isotropic configurations, the imaginary part for a large number of sums vanishes. Hence, the accuracy of ${\mathcal Re}X_q$ and ${\mathcal Im}X_q$ is explained.

Besides calculating the general accuracy of the model, one can also analyse the {\it confusion matrix} that contains all the classes in both the horizontal and vertical directions. The element of the matrix at $(i, j)$ tells us how many samples belonging to class $C_i$ have been assigned to class $C_j$ by the algorithm. Hence, the leading diagonal contains all the correct predictions. All presented confusion matrices were generated using training:testing ratio 25:75 and the 3-fold cross-validation procedure.

\begin{figure}[ht]
\centering
\includegraphics[clip, trim=0mm 5mm 3mm 0mm, width=.5491\textwidth]{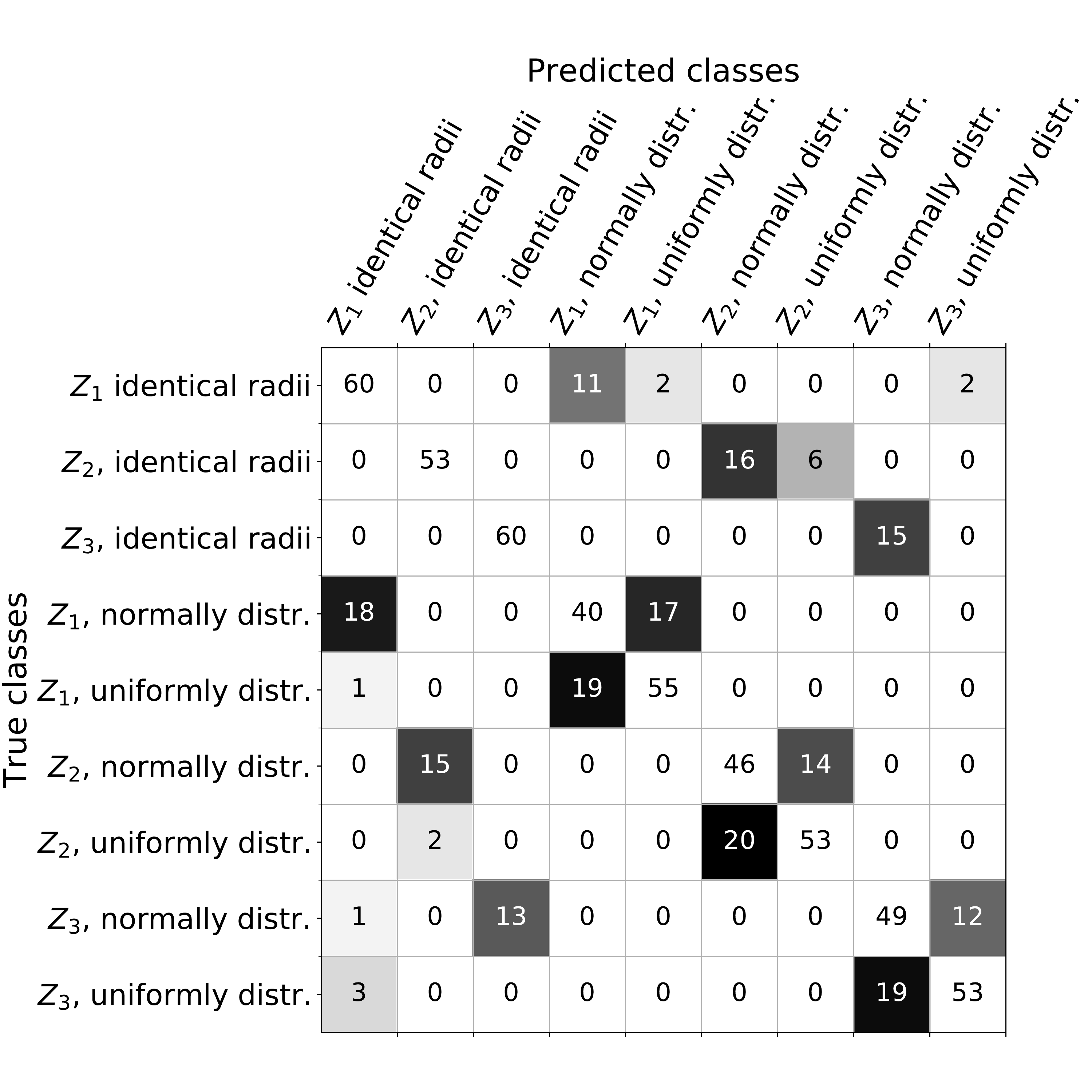} 
\includegraphics[clip, trim=89mm 5mm 0mm 0mm, width=.39235\textwidth]{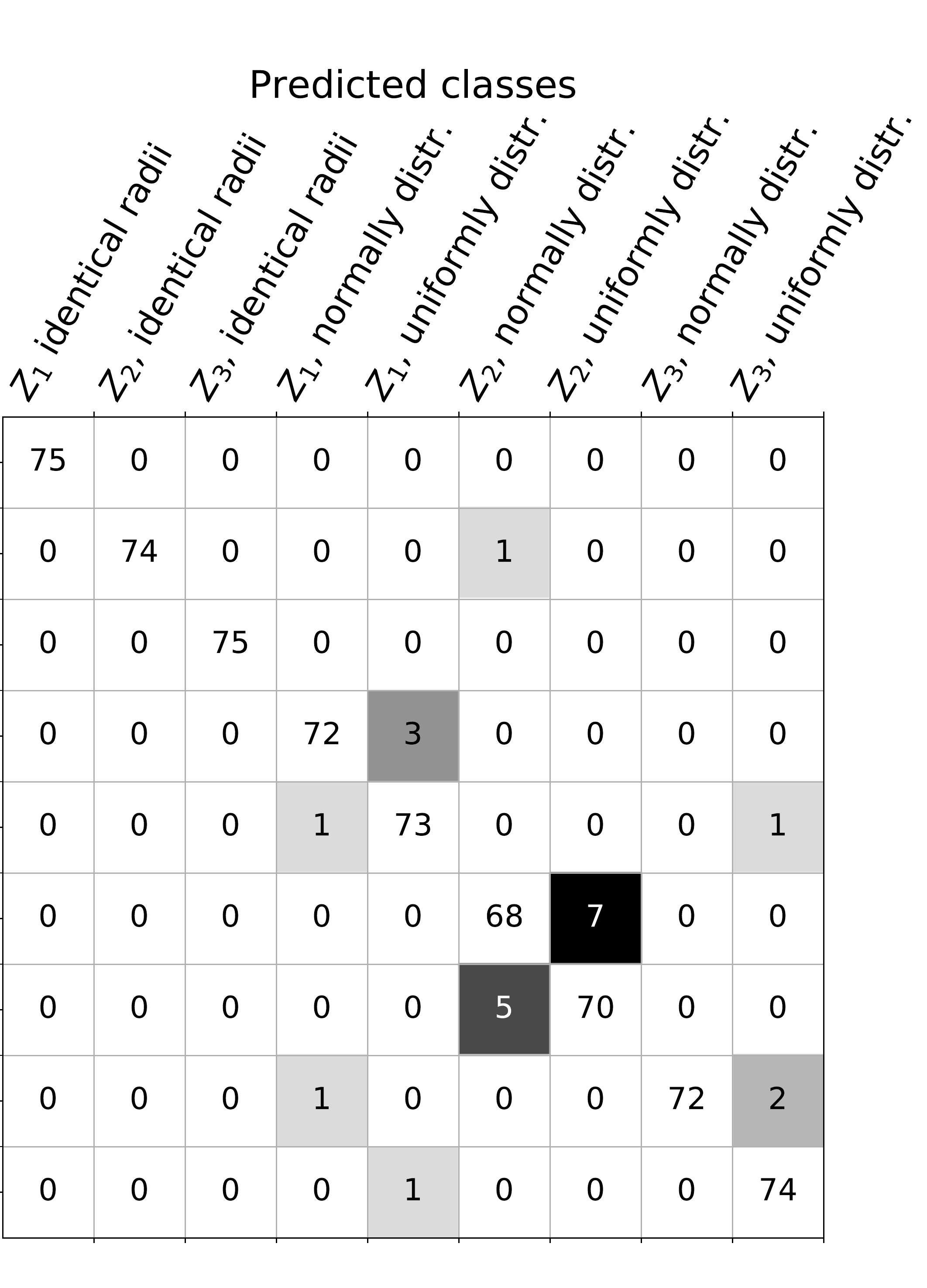}
\caption{Confusion matrices for models built on the vector $|X_q|$ for $q=5$ (left) and $q=10$ (right).} 
\label{fig:conf_matr_disks_abs}
\end{figure}

\begin{figure}[ht]
\centering
\includegraphics[clip, trim=0mm 5mm 3mm 0mm, width=.578\textwidth]{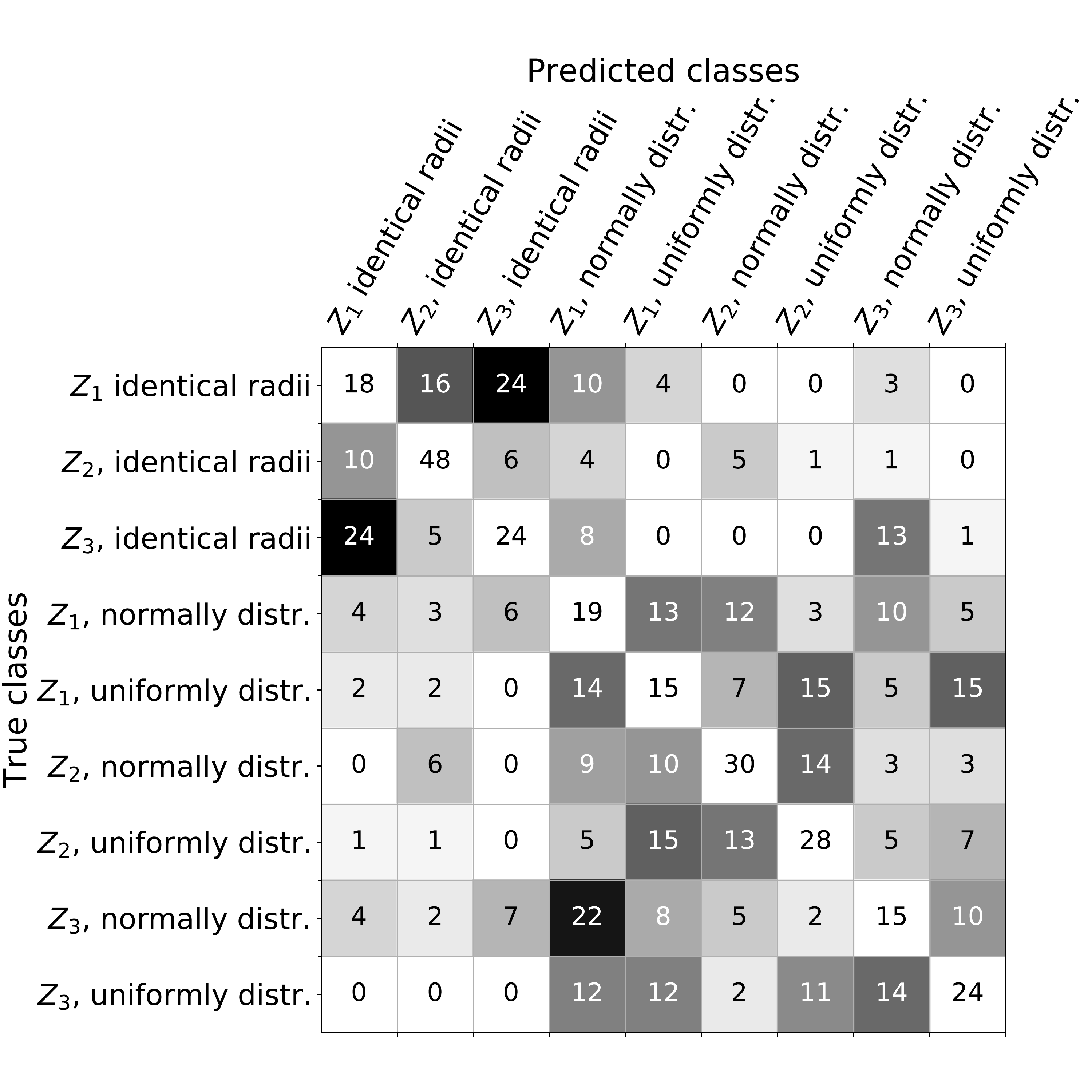}
\caption{Confusion matrix for model built on the vector ${\mathcal Arg}X_q$ for $q=10$.} 
\label{fig:conf_matr_disks_arg}
\end{figure}

Let us analyse confusion matrix for the model built on the vector $|X_q|$ for $q=5$ (see Fig.~\ref{fig:conf_matr_disks_abs} (left)). First of all, algorithm almost always assigns proper variable $Z_j$ ($j=1,2,3$). Secondly, sample belonging to the class with identical radii is never confused with another identical radii class (note the white $3\times 3$ square in the top-left corner of the matrix), rather with a polydispersed one with normally distributed radii and the same $Z_k$ (first three rows of the matrix). Finally, another source of errors is to confuse uniformly distributed radii with normally  distributed ones among the same $Z_k$ (see black spots around the leading diagonal of the $6\times 6$ square in the bottom-right corner of the matrix).
As the order $q$ grows, the accuracy of the model increases, see the confusion matrix for $q=10$ (Fig.~\ref{fig:conf_matr_disks_abs} (right)). One can see that there is still a small problem with confusing radii distribution among the $Z_2$ class.
Fig.~\ref{fig:conf_matr_disks_arg} shows the confusion matrix for the vector ${\mathcal Arg}X_q$ for $q=10$. The matrix reflects sources of the poor performance of ${\mathcal Arg}X_q$.

\subsection{Non-circular inclusions}
\label{sect:shapes}



\begin{figure}[!htb]
\centering
\begin{tabular}{cccc}

\includegraphics[clip, trim=29mm 25mm 20mm 15mm, width=.15\textwidth]{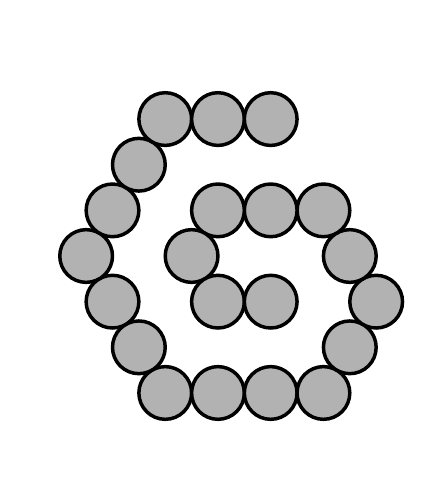}0&
\includegraphics[clip, trim=29mm 25mm 20mm 15mm, width=.15\textwidth]{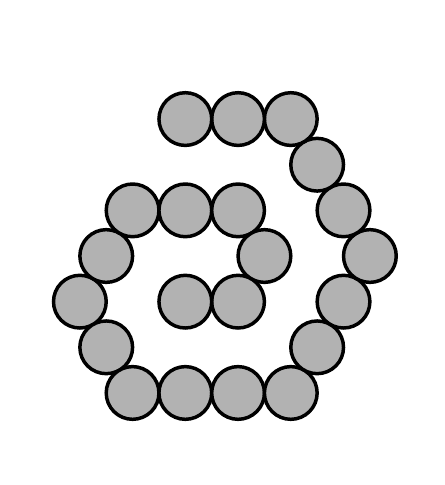}1
&
\includegraphics[clip, trim=14mm 10mm 5mm 0mm, width=.20\textwidth]{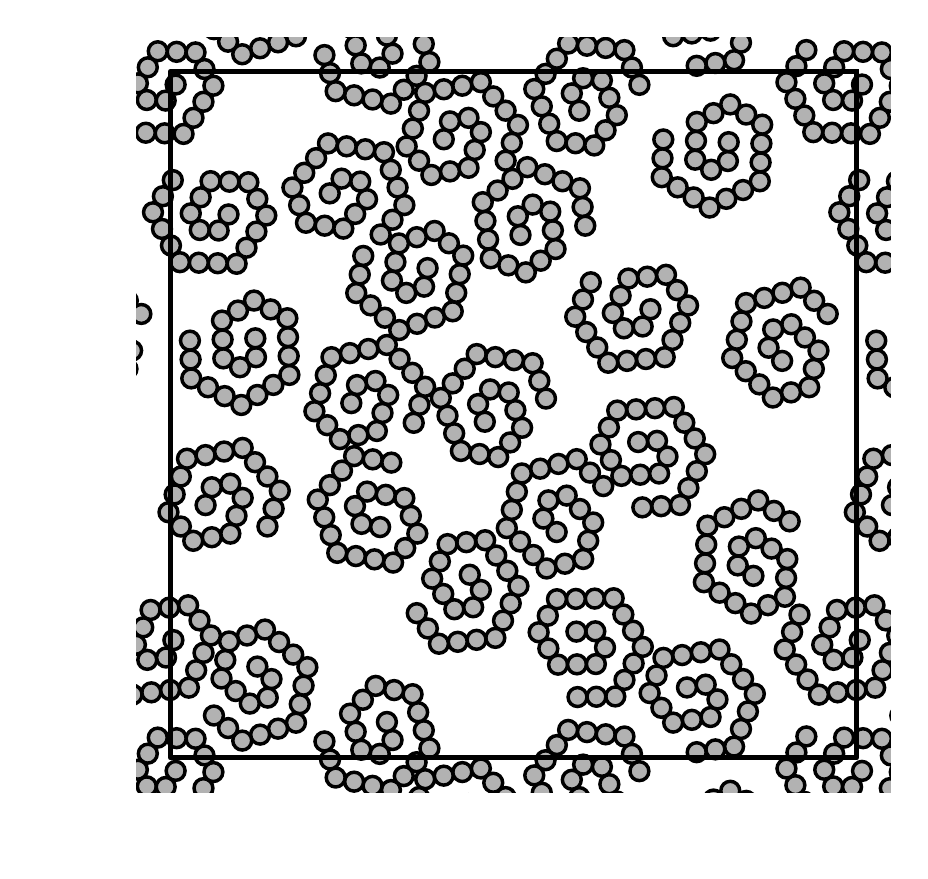}
&
\includegraphics[clip, trim=14mm 10mm 5mm 0mm, width=.20\textwidth]{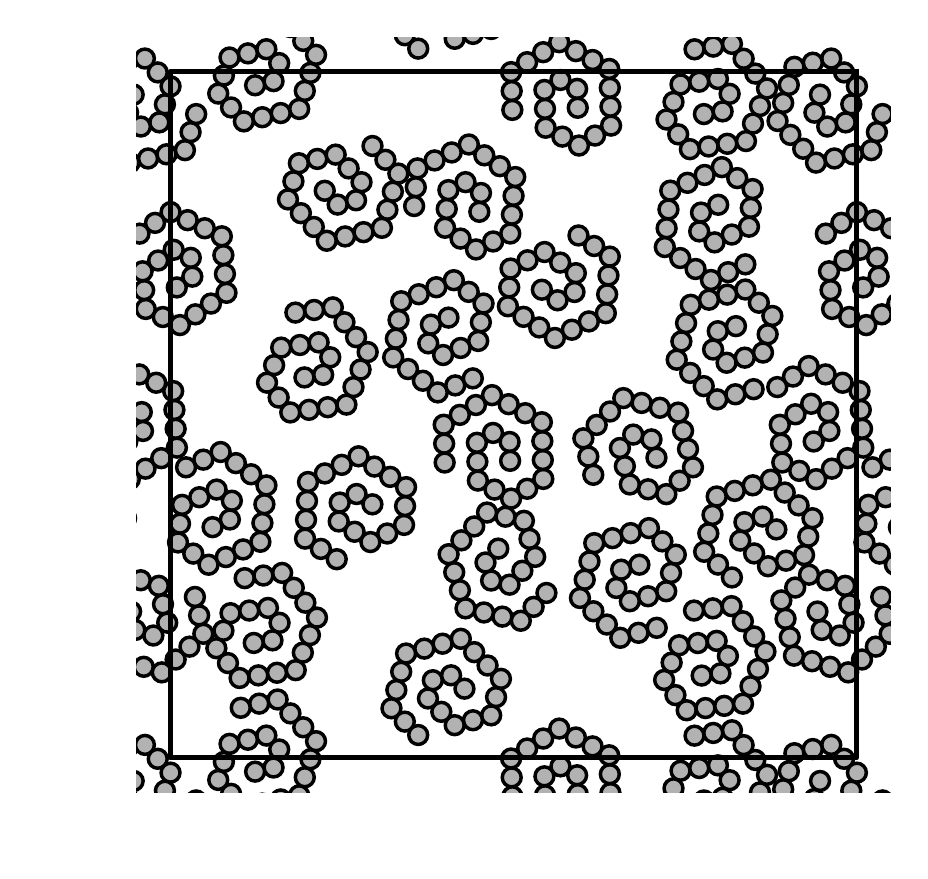}
\\

\includegraphics[clip, trim=29mm 25mm 20mm 15mm, width=.15\textwidth]{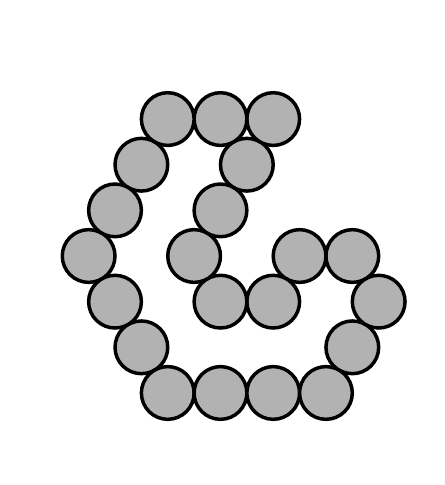}2
&
\includegraphics[clip, trim=29mm 25mm 20mm 15mm, width=.15\textwidth]{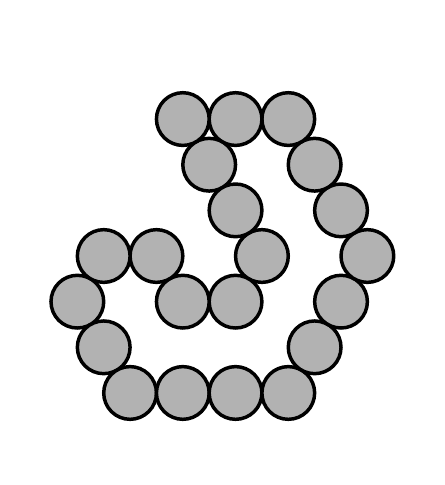}3
&
\includegraphics[clip, trim=14mm 10mm 5mm 0mm, width=.20\textwidth]{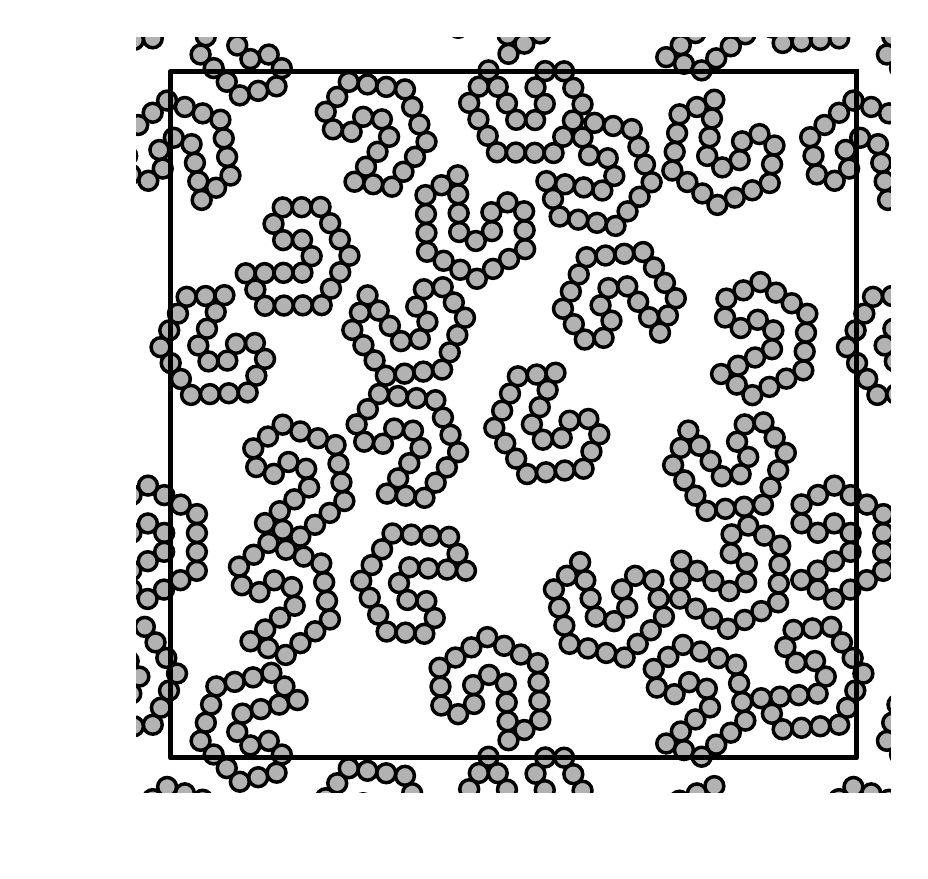}
&
\includegraphics[clip, trim=14mm 10mm 5mm 0mm, width=.20\textwidth]{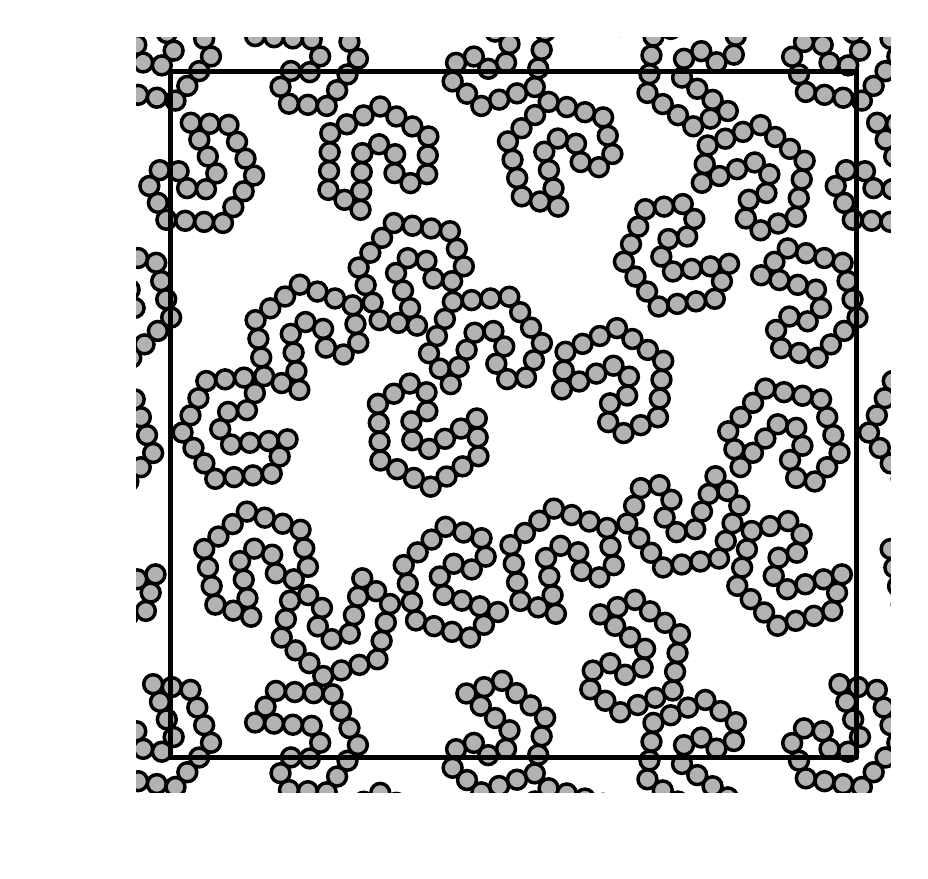}
\\

\includegraphics[clip, trim=29mm 25mm 20mm 15mm, width=.15\textwidth]{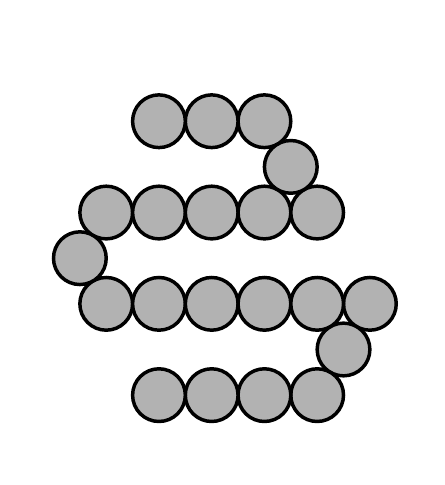}4
&
\includegraphics[clip, trim=29mm 25mm 20mm 15mm, width=.15\textwidth]{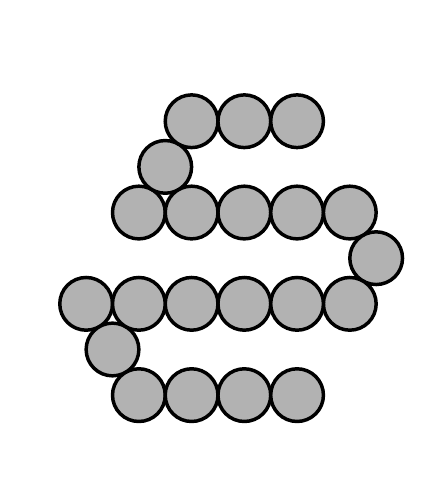}5
&
\includegraphics[clip, trim=14mm 10mm 5mm 0mm, width=.20\textwidth]{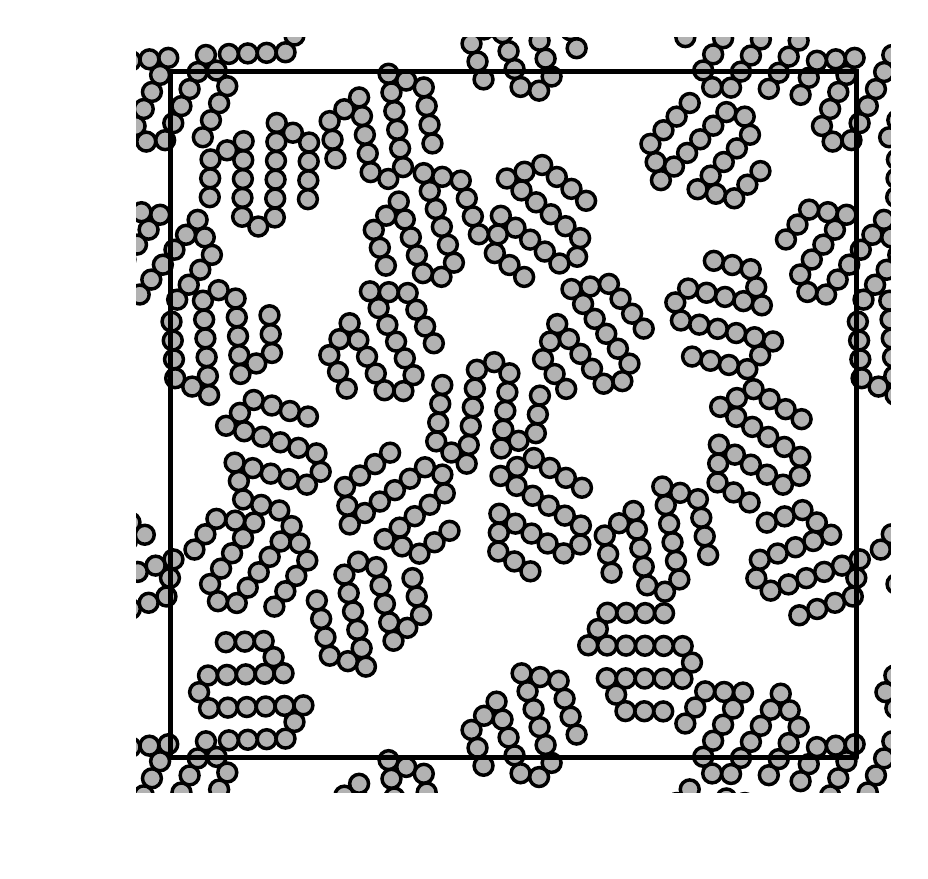}
&
\includegraphics[clip, trim=14mm 10mm 5mm 0mm, width=.20\textwidth]{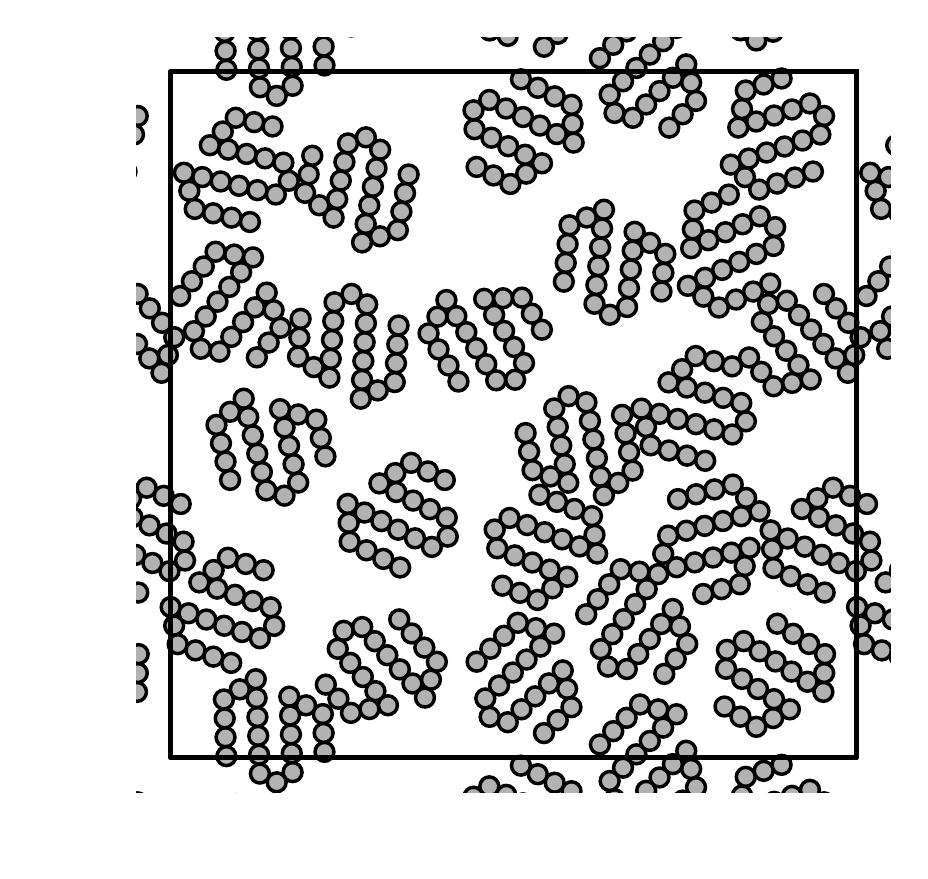}
\\

\includegraphics[clip, trim=29mm 25mm 20mm 15mm, width=.15\textwidth]{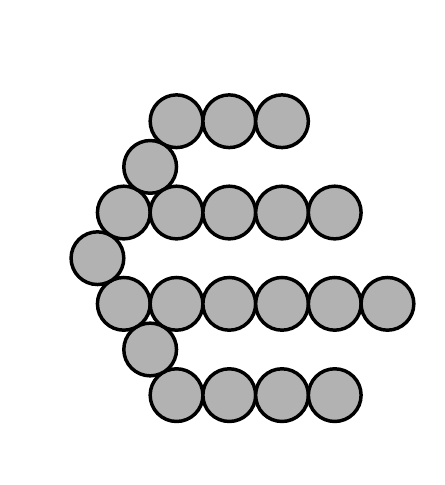}6
&
\includegraphics[clip, trim=29mm 25mm 20mm 15mm, width=.15\textwidth]{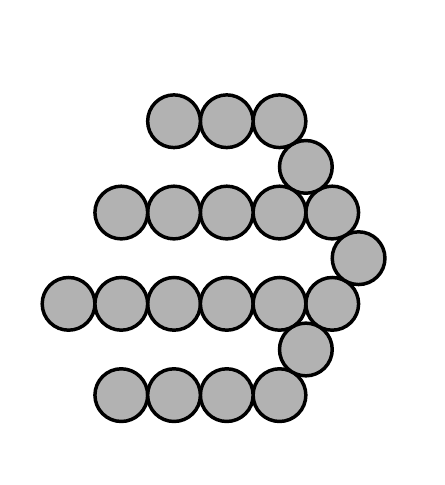}7
&
\includegraphics[clip, trim=14mm 10mm 5mm 0mm, width=.20\textwidth]{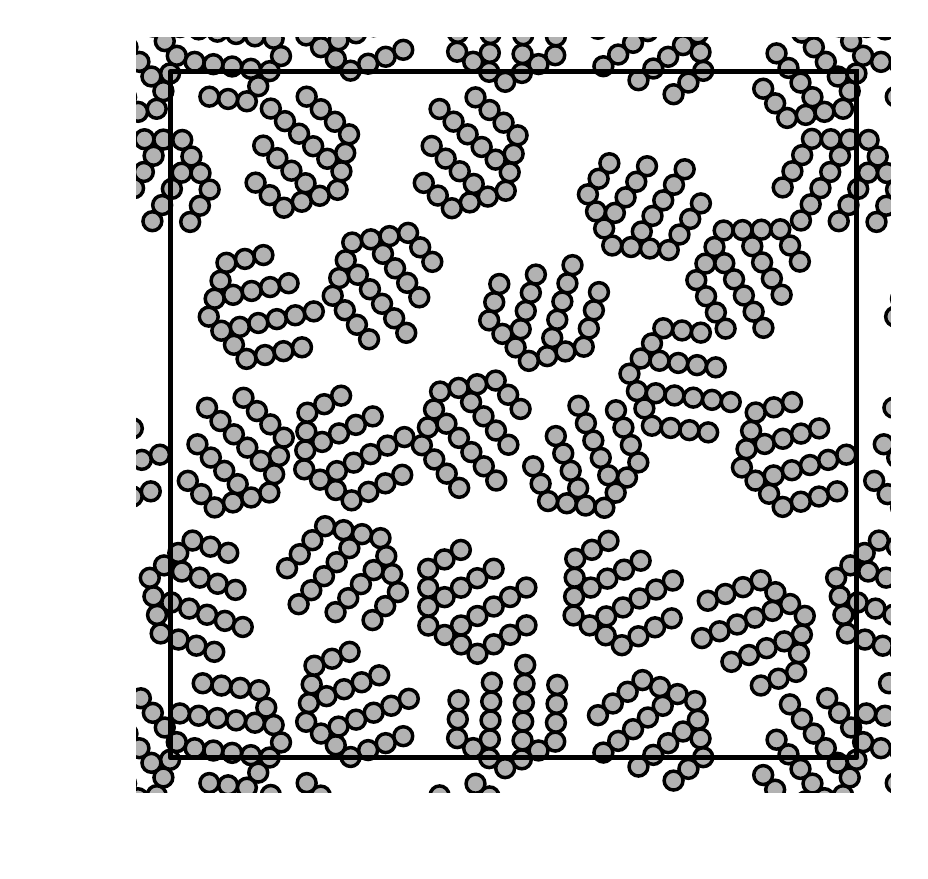}
&
\includegraphics[clip, trim=14mm 10mm 5mm 0mm, width=.20\textwidth]{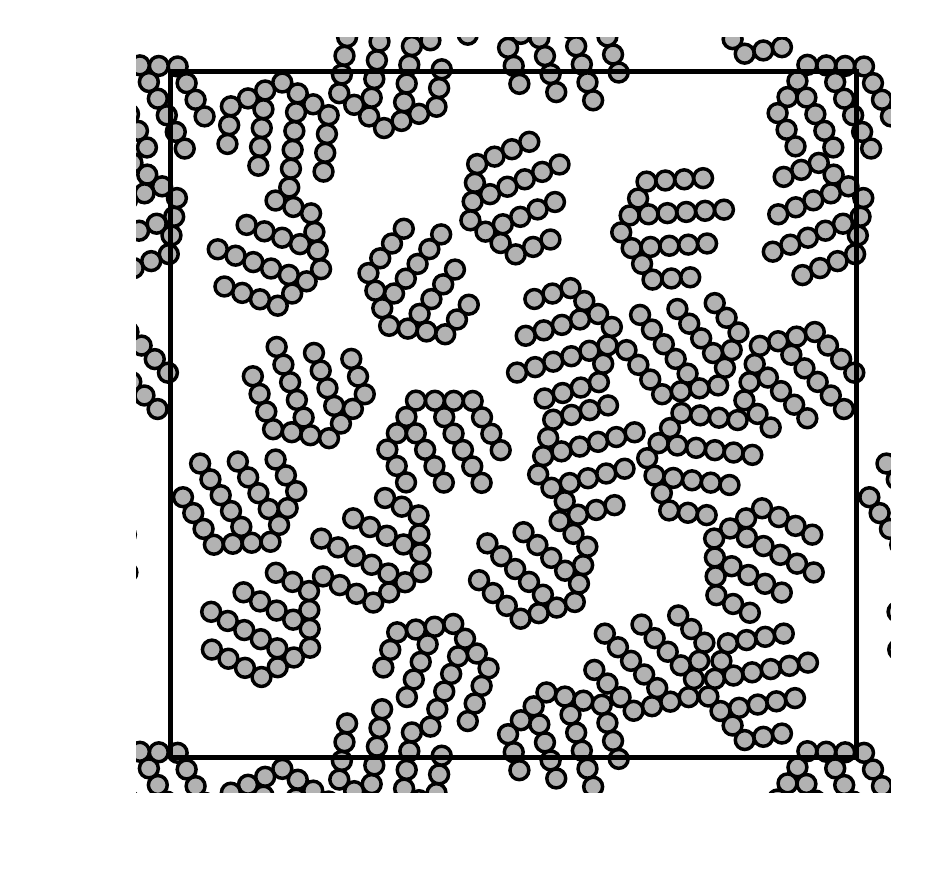}
\\

\includegraphics[clip, trim=29mm 25mm 20mm 15mm, width=.15\textwidth]{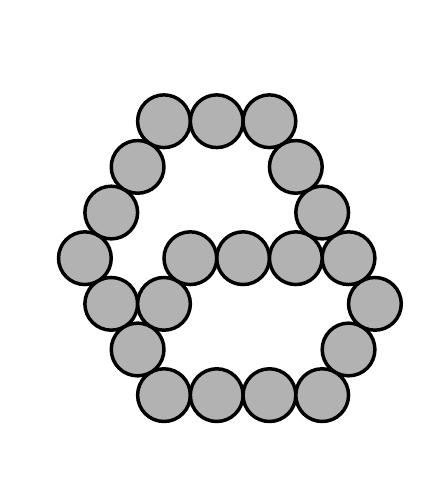}8
&
\includegraphics[clip, trim=29mm 25mm 20mm 15mm, width=.15\textwidth]{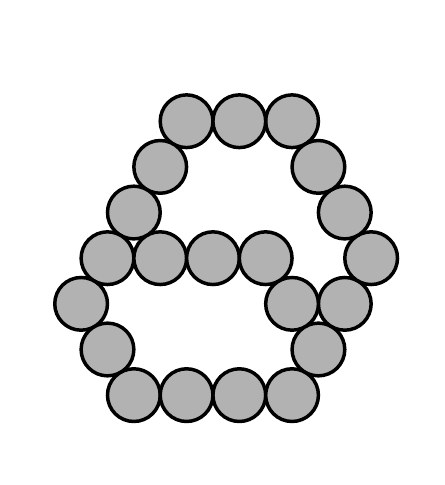}9
&
\includegraphics[clip, trim=14mm 10mm 5mm 0mm, width=.20\textwidth]{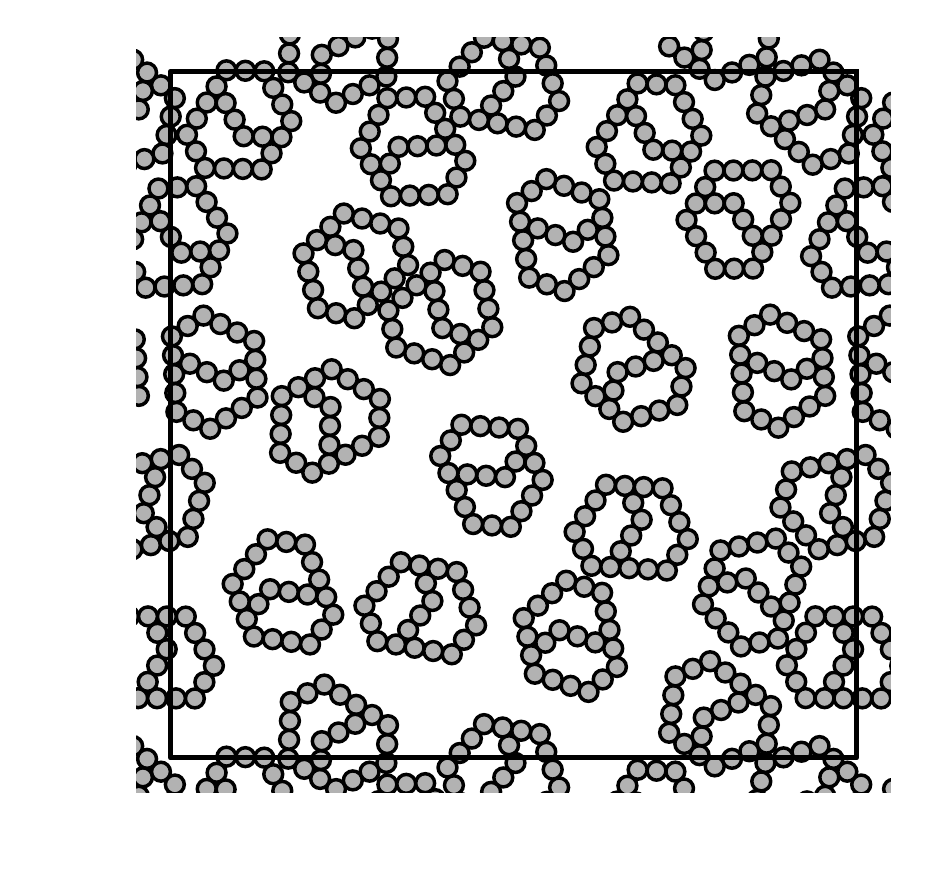}
&
\includegraphics[clip, trim=14mm 10mm 5mm 0mm, width=.20\textwidth]{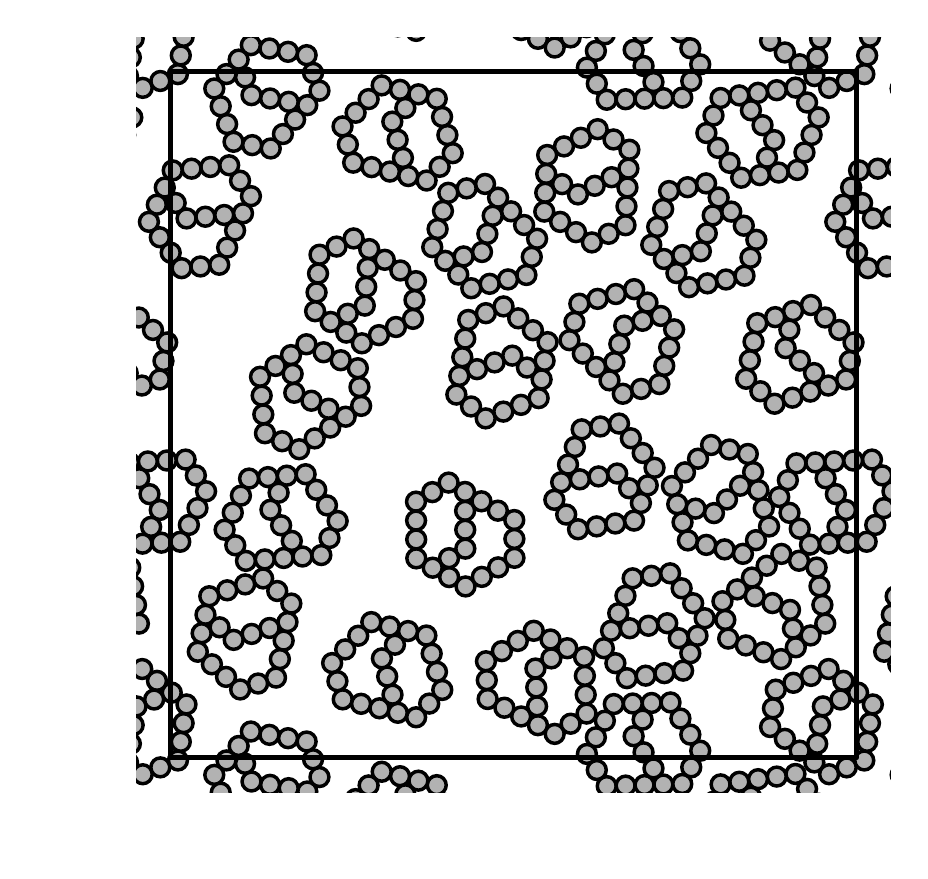}
\\

\end{tabular}
\caption{Similar shapes consisting of the same number of disks of fixed concentration (left) and corresponding samples of distributions generated via the RSA protocol (right).}
\label{fig:distributions_shapes}
\end{figure}

In the present section we demonstrate shape-sensitivity of structural sums. Inclusions of non-circular shape can be approximated by clusters of disks. Consider similar shapes shown in Fig.~\ref{fig:distributions_shapes}, consisting of 21 identical disks each. In order to generate samples of distributions, we apply the RSA protocol for each shape. Since the chosen protocol is the same for each shape, one can assume that the potential differences between classes will be caused by the shape of inclusions, rather than their planar distribution.
Simulated samples are configurations of concentration $\nu=0.3$ (see examples in Fig.~\ref{fig:distributions_shapes}).

We generated a set of 100 samples for each class of distributions and adopted the same classification scheme as in section~\ref{sect:disks}. The results, in varying both the size of training data and the order of modified feature vector~$X_q$, are presented in Table~\ref{tab:acc_shapes}.

One can see that application of vectors $|X_q|$ and ${\mathcal Re}X_q$, which performed well in the preceding example, yields poor results. This clearly shows that the methods based only on distances between centres of disks may be insufficient in some cases. In contrast to the preceding experiment, very good results are obtained for a relatively small training set (4\%-16\%  of the dataset) for vectors ${\mathcal Im}X_q$ and ${\mathcal Arg}X_q$ of orders 6-10. It is worth noting that the analysis of distances is equivalent to analysis based on the autocorrelation (2-point correlation) function. Also note that a purely random classifier has the expected accuracy equal 0.1.

\begin{table}[!t]
\caption{Accuracy of classification of distributions of shapes for $|X_q|$, ${\mathcal Re}X_q$, ${\mathcal Im}X_q$, and ${\mathcal Arg}X_q$.}
\label{tab:acc_shapes}
\setlength{\tabcolsep}{3pt}
\centering\scriptsize
\begin{tabular}{cccccccccccc}
\cline{1-12}
\multicolumn{2}{c}{train}  & \multicolumn{10}{c}{order of feature vector} \\ \cline{3-12}

\multicolumn{2}{c}{size} &        1  &        2  &        3  &        4  &        5  &        6  &        7  &        8  &        9  &        10 \\   \hline 

\multicolumn{12}{l}{$|X_q|$}\\
&16\%         &  0.135 &  0.279 &  0.505 &  0.504 &  0.503 &  0.505 &  0.500 &  0.507 &  0.505 &  0.502 \\
&28\%         &  0.133 &  0.277 &  0.502 &  0.499 &  0.493 &  0.499 &  0.497 &  0.499 &  0.497 &  0.498 \\
&40\%         &  0.136 &  0.278 &  0.513 &  0.511 &  0.495 &  0.502 &  0.499 &  0.499 &  0.496 &  0.501 \\

\multicolumn{12}{l}{${\mathcal Re}X_q$}\\
&16\%         &  0.137 &  0.278 &  0.505 &  0.503 &  0.501 &  0.502 &  0.502 &  0.508 &  0.505 &  0.498 \\
&28\%         &  0.133 &  0.278 &  0.501 &  0.499 &  0.495 &  0.498 &  0.500 &  0.501 &  0.496 &  0.498 \\
&40\%         &  0.135 &  0.278 &  0.515 &  0.510 &  0.501 &  0.503 &  0.505 &  0.499 &  0.502 &  0.502 \\

\multicolumn{12}{l}{${\mathcal Im}X_q$}\\
& 4\%         &  0.137 &  0.137 &  0.131 &  0.142 &  0.426 &  0.621 &  0.748 &  0.770 &  0.871 &  0.924 \\
& 7\%         &  0.143 &  0.143 &  0.138 &  0.145 &  0.468 &  0.700 &  0.852 &  0.890 &  0.958 &  0.976 \\
&10\%         &  0.142 &  0.142 &  0.140 &  0.150 &  0.491 &  0.728 &  0.886 &  0.928 &  0.981 &  0.988 \\
&13\%         &  0.142 &  0.142 &  0.143 &  0.157 &  0.497 &  0.752 &  0.909 &  0.943 &  0.989 &  0.994 \\
&16\%         &  0.140 &  0.140 &  0.144 &  0.154 &  0.500 &  0.750 &  0.910 &  0.943 &  0.987 &  0.990 \\
&19\%         &  0.141 &  0.141 &  0.144 &  0.160 &  0.513 &  0.777 &  0.919 &  0.947 &  0.991 &  0.995 \\
&22\%         &  0.141 &  0.141 &  0.146 &  0.159 &  0.510 &  0.776 &  0.918 &  0.945 &  0.990 &  0.994 \\
&25\%         &  0.139 &  0.139 &  0.147 &  0.155 &  0.517 &  0.780 &  0.931 &  0.961 &  0.994 &  0.995 \\
&28\%         &  0.142 &  0.142 &  0.146 &  0.159 &  0.519 &  0.782 &  0.935 &  0.966 &  0.997 &  0.998 \\
&31\%         &  0.139 &  0.139 &  0.143 &  0.160 &  0.526 &  0.791 &  0.927 &  0.956 &  0.995 &  0.997 \\
&34\%         &  0.147 &  0.147 &  0.151 &  0.167 &  0.525 &  0.803 &  0.932 &  0.962 &  0.996 &  0.996 \\
&37\%         &  0.146 &  0.146 &  0.150 &  0.159 &  0.530 &  0.806 &  0.947 &  0.974 &  0.998 &  0.998 \\
&40\%         &  0.142 &  0.142 &  0.150 &  0.162 &  0.534 &  0.804 &  0.936 &  0.963 &  0.995 &  0.996 \\

\multicolumn{12}{l}{${\mathcal Arg}X_q$}\\
& 4\%         &  0.139 &  0.139 &  0.123 &  0.113 &  0.288 &  0.457 &  0.469 &  0.413 &  0.309 &  0.305 \\
& 7\%         &  0.141 &  0.141 &  0.131 &  0.130 &  0.370 &  0.609 &  0.721 &  0.673 &  0.718 &  0.693 \\
&10\%         &  0.141 &  0.141 &  0.139 &  0.137 &  0.419 &  0.675 &  0.814 &  0.799 &  0.817 &  0.789 \\
&13\%         &  0.144 &  0.144 &  0.143 &  0.139 &  0.426 &  0.688 &  0.824 &  0.857 &  0.871 &  0.821 \\
&16\%         &  0.139 &  0.139 &  0.143 &  0.140 &  0.433 &  0.712 &  0.830 &  0.880 &  0.878 &  0.840 \\
&19\%         &  0.140 &  0.140 &  0.138 &  0.138 &  0.433 &  0.710 &  0.839 &  0.895 &  0.907 &  0.881 \\
&22\%         &  0.142 &  0.142 &  0.141 &  0.139 &  0.423 &  0.709 &  0.838 &  0.898 &  0.906 &  0.881 \\
&25\%         &  0.139 &  0.139 &  0.139 &  0.141 &  0.441 &  0.737 &  0.858 &  0.914 &  0.920 &  0.891 \\
&28\%         &  0.141 &  0.141 &  0.145 &  0.143 &  0.447 &  0.736 &  0.850 &  0.917 &  0.915 &  0.890 \\
&31\%         &  0.139 &  0.139 &  0.140 &  0.137 &  0.448 &  0.730 &  0.844 &  0.930 &  0.932 &  0.899 \\
&34\%         &  0.144 &  0.144 &  0.155 &  0.144 &  0.454 &  0.740 &  0.857 &  0.938 &  0.941 &  0.924 \\
&37\%         &  0.147 &  0.147 &  0.141 &  0.138 &  0.443 &  0.743 &  0.860 &  0.937 &  0.935 &  0.923 \\
&40\%         &  0.143 &  0.143 &  0.147 &  0.133 &  0.453 &  0.740 &  0.853 &  0.931 &  0.941 &  0.933 \\

\end{tabular}
\end{table}

\begin{figure}[ht]
\centering
\includegraphics[clip, trim=0mm 0mm 0mm 0mm, width=.55\textwidth]{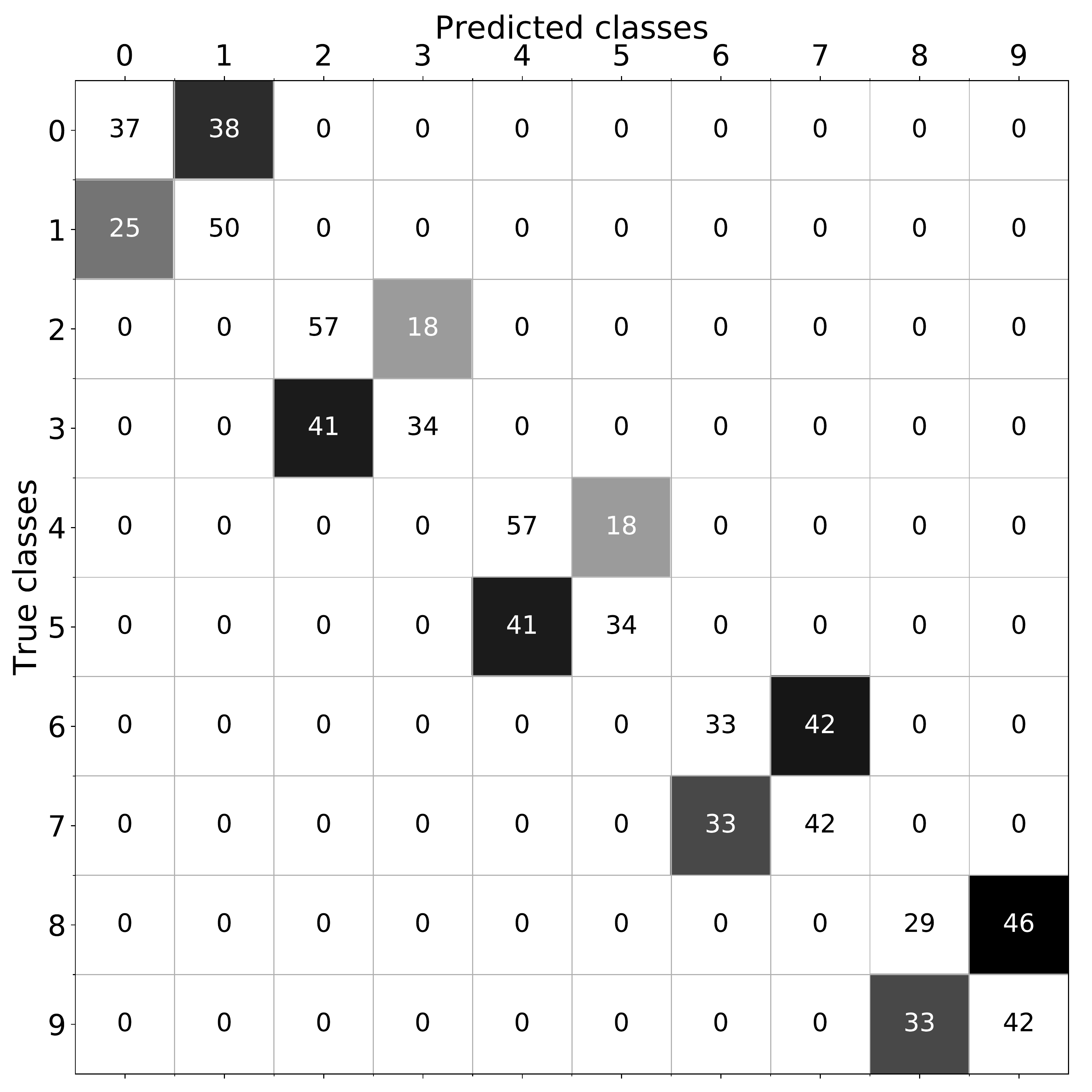} 
\caption{Confusion matrix for model built on the vector $|X_q|$ for $q=3$. Numbers of classes are consistent with the numbering on Fig.~\ref{fig:distributions_shapes}.} 
\label{fig:conf_matr_shapes}
\end{figure}

Let us find out what are the sources of the poor performance of $|X_q|$. In order to do that, we analyse the confusion matrix for $q=3$ (Fig.~\ref{fig:conf_matr_shapes}). Black spots around the leading diagonal tell us that its main difficulty is to distinguish between mirrored shapes. Otherwise, it performs surprisingly well, considering its small size.

\section{Application: Irregularity of random structures}
\label{sect:applications}

\begin{figure}[ht]
\centering
\includegraphics[clip, trim=10mm 5mm 10mm 0mm, width=.54\textwidth]{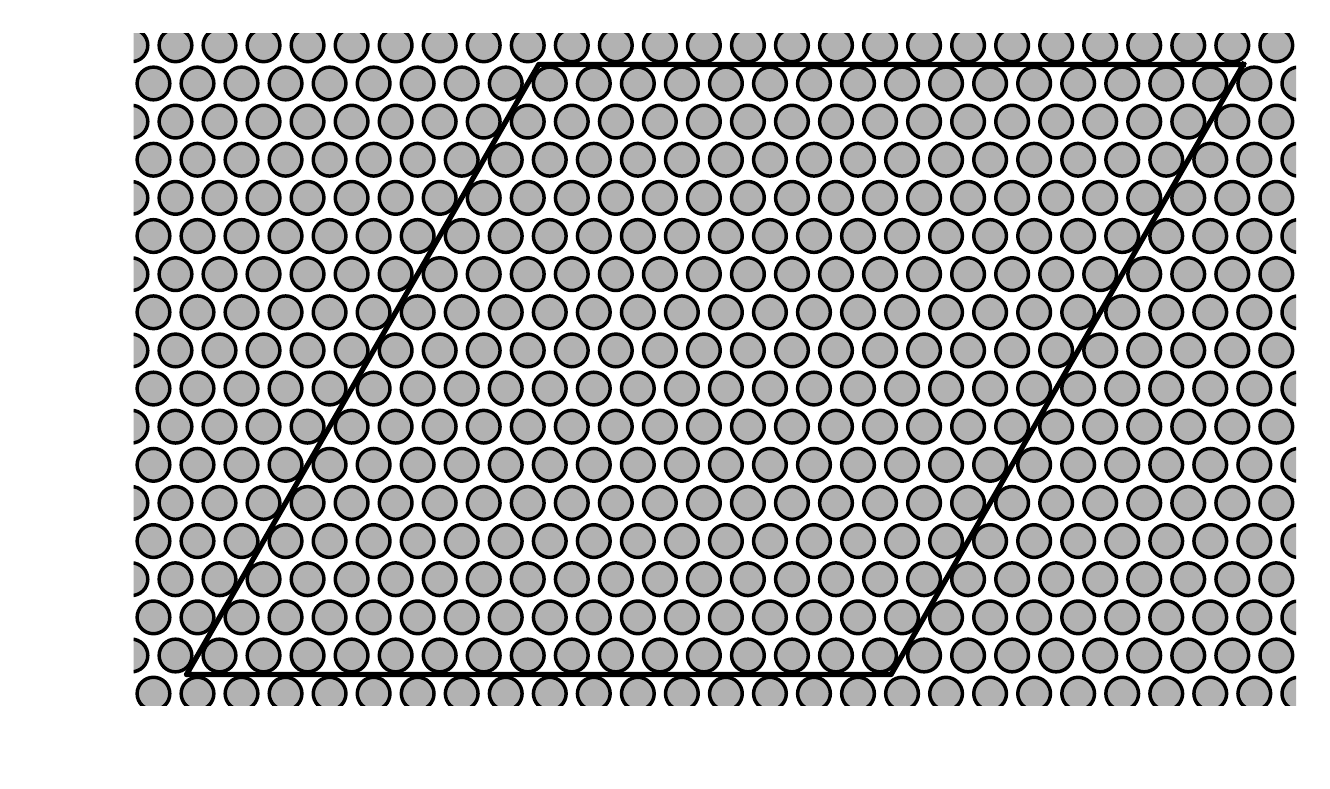} 
\includegraphics[clip, trim=10mm 5mm 0mm 0mm, width=.351\textwidth]{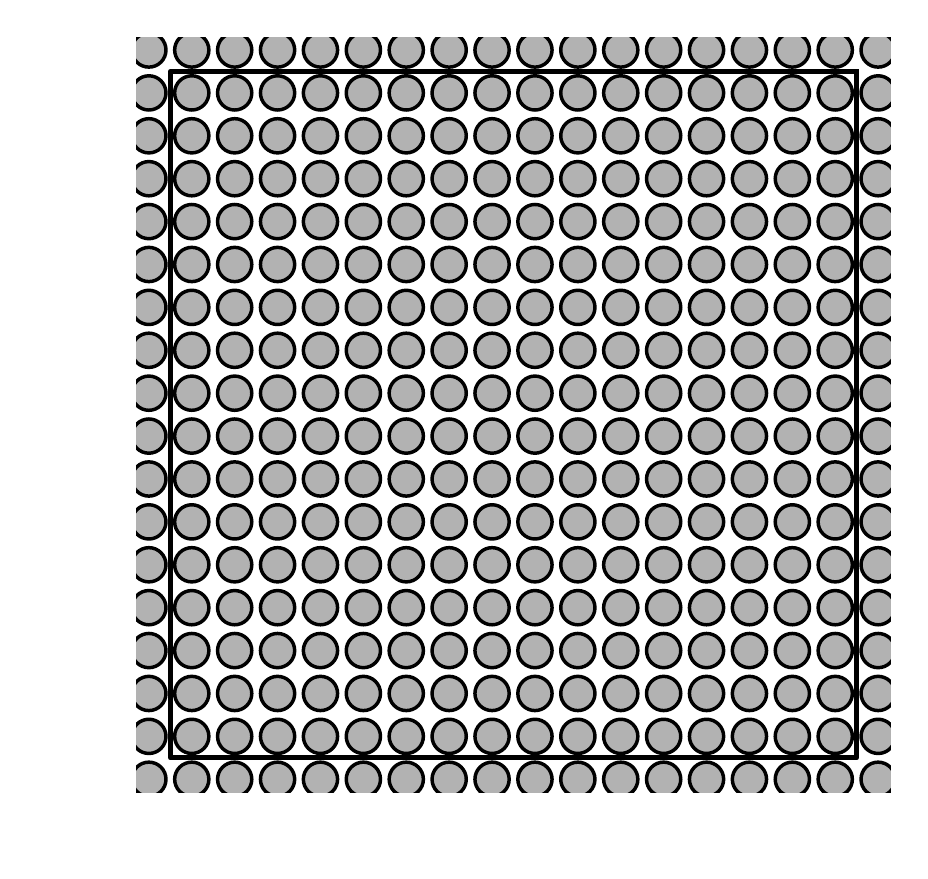} 
\caption{Regular arrays of identical disks: hexagonal array (left) and square array (right).} 
\label{fig:reg_hex}
\end{figure}

It is known from the theory \cite{MitRyl2012} that structural sums attain their extrema for optimally distributed disks. For example, in case of identical disks the regular hexagonal array (see Fig.~\ref{fig:reg_hex}) allows to achieve the maximal concentration. All distributions considered in section~\ref{sect:disks} have the same concentration $\nu=0.5$, however each of them shows different level of {\it irregularity}. For example, the sample in the second column and the first row ($Z_2$, identical radii) seems to be the most regular among the others. On the other hand, the pattern in the third column and the second row ($Z_3$, normally distributed radii) looks very irregular. Let us apply structural sums in order to express this property quantitatively.
For this purpose, we will present our data in two-dimensional space. In order to do this, let us apply the following simple {\it exhaustive} feature selection scheme. 
We train Naive Bayes model for only two features, considering all pair combinations of sums from $|X_{10}|$. Moreover, we consider only structural sums of the form $e_{p,p}$ ($2\leq p\leq 10$). The experiments are performed using training:testing ratio 25:75 and the 3-fold cross-validation procedure. The average of resulting accuracy scores is assigned to each pair of features. Table~\ref{tab:2features_abs} shows pairs with the accuracy score greater than 0.95. Let us select sums of lowest orders for further considerations, namely $e_{3,3}$ and $e_{8,8}$.

\begin{table}[!htb]
\caption{Mean accuracy of classification of distributions of disks. The accuracy is computed as the average of the values computed via 3-fold cross-validation procedure using Gaussian Naive Bayes classification algorithm.}
\label{tab:2features_abs}
\setlength{\tabcolsep}{3pt}
\centering
\begin{tabular}{ccclccccl}
\hline
accuracy  &&  \multicolumn{2}{c}{features} && accuracy  &&  \multicolumn{2}{c}{features}    \\

\hline
0.987 && $e_{4, 4}$ & $e_{10, 10}$ & & 0.973 && $e_{5, 5}$ & $e_{8, 8}$\\
0.987 && $e_{3, 3}$ & $e_{10, 10}$ & & 0.972 && $e_{5, 5}$ & $e_{9, 9}$\\
0.977 && $e_{5, 5}$ & $e_{10, 10}$ & & 0.969 && $e_{3, 3}$ & $e_{8, 8}$\\
0.977 && $e_{4, 4}$ & $e_{9, 9}$ & & 0.968 && $e_{3, 3}$ & $e_{9, 9}$\\
0.973 && $e_{4, 4}$ & $e_{8, 8}$ & & 0.951 && $e_{6, 6}$ & $e_{10, 10}$\\
\end{tabular}
\end{table}

\begin{figure}[!ht]
\centering
\includegraphics[clip, trim=0mm 6mm 0mm 6mm, width=.75\textwidth]{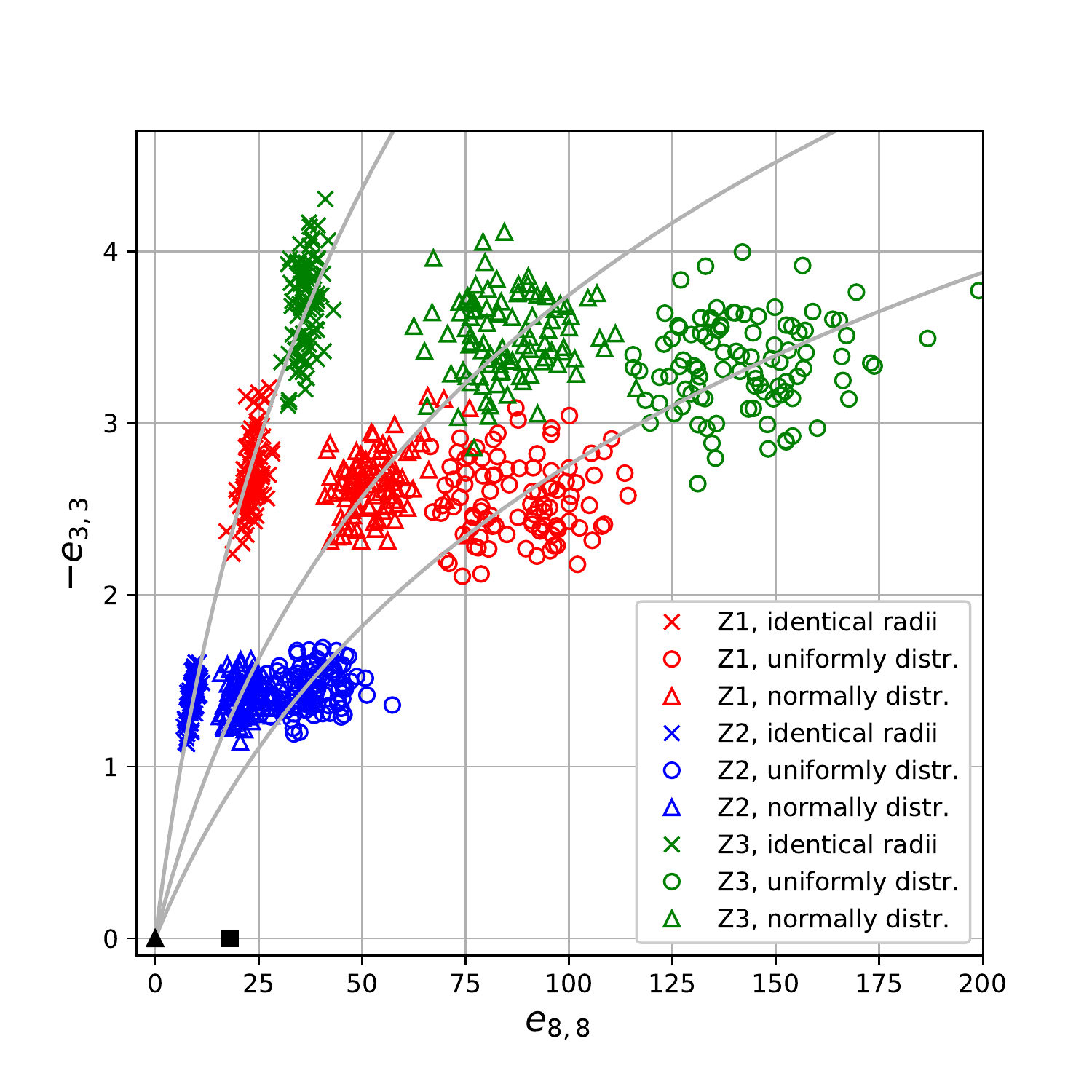} 
\caption{Values of $-e_{3,3}$ against $e_{8,8}$ for samples from considered distributions. The fitted curves are 
$3.118\log(0.061 x +1)$ (identical radii, crosses), 
$2.526\log(0.034 x +1)$ (normally distributed radii, triangles),
$1.987\log(0.028 x +1)$ (uniformly distributed radii, disks).  
Black triangle and square are for hexagonal and square regular arrays, respectively.
} 
\label{fig:2d_features}
\end{figure}

 Consider the hexagonal array of disks as the most regular one. In such a case both minimum of $e_{8,8}$ and the maximum of $e_{3, 3}$ are equal zero. Intuitively, the more regular system of disks is, the closer structural sums are to their extrema. Fig.~\ref{fig:2d_features}, showing the plot of $-e_{3,3}$ against $e_{8,8}$, confirms this intuitive remark. One can also observe the convergence of the regularity of distributions to the hexagonal case. It occurs that the data can be fitted to a simple logarithmic model $a\log(b x +1)$ passing through the origin (see Fig.~\ref{fig:2d_features}). The curves illustrate paths of convergence of the sums and also represent each distribution $Z_j$ ($j=1,2,3$).

\begin{figure}[ht]
\centering
\includegraphics[clip, trim=2mm 6mm 10mm 6mm, width=.75\textwidth]{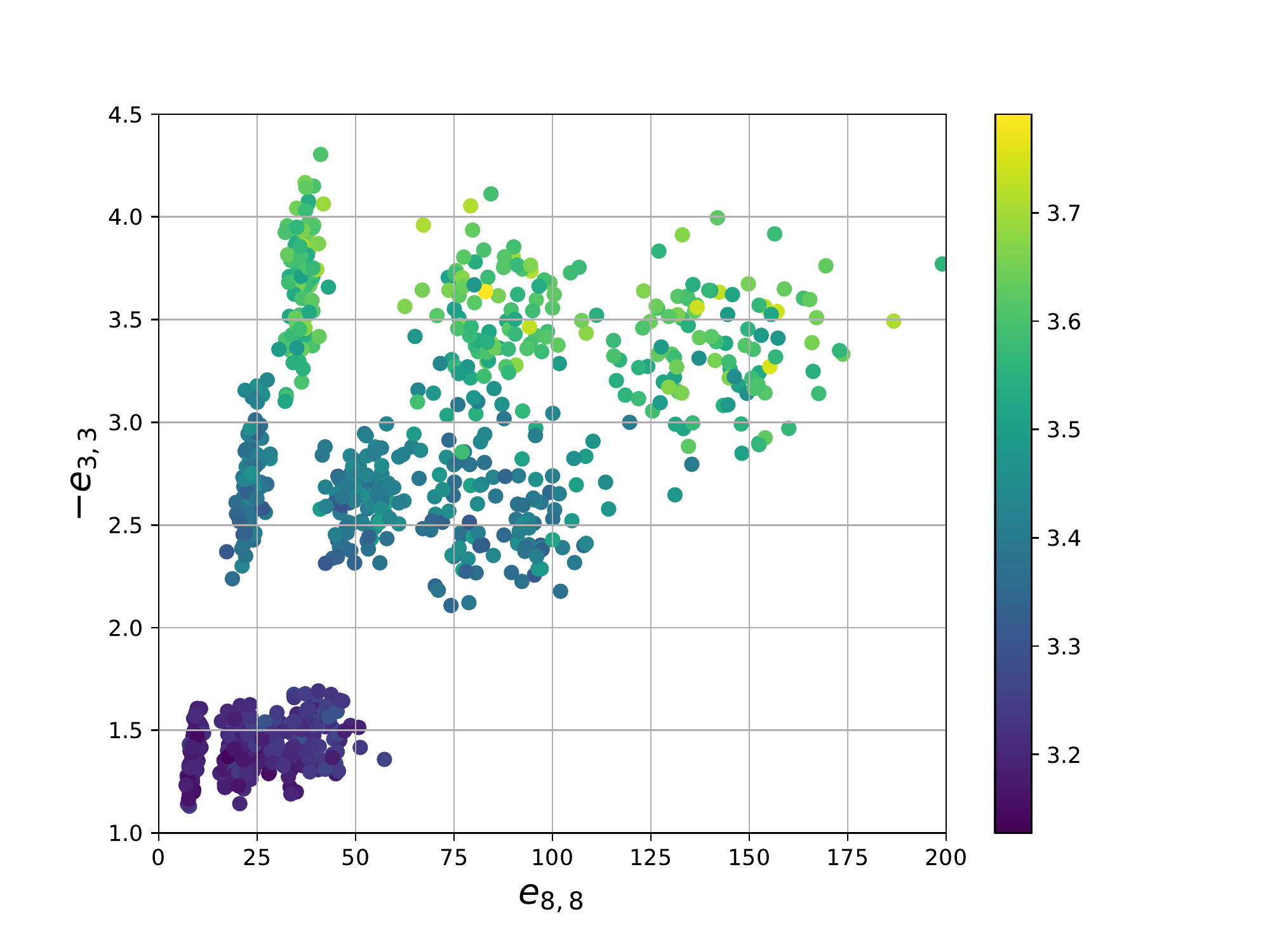} 
\caption{Values of $-e_{3,3}$ against $e_{8,8}$ for samples from considered distributions. The colors represent the values of the effective conductivity of the composite modelled by a given sample.
} 
\label{fig:EC_color}
\end{figure}

Let us now attempt to describe geometric meaning of selected sums. 
One can observe that $e_{8,8}$ is related to the heterogeneity of disks in a pattern. Moreover, $e_{3,3}$ seems to be reflecting {\it clustering} of disks. This can be explained on the basis of the effective conductivity of composites modelled by considered samples. The feature vector $X_q$ provides components for an approximation of the effective conductivity (EC) formula (for more details, see section~\ref{sect:conductivity} as well as \cite{GluMitNaw2017Book}). The values of EC are presented in Fig.~\ref{fig:EC_color}. Since all samples have the same concentrations of inclusions, hence the only possible source of differences in EC lies in the geometry of a system. One can see that the higher values of EC are common in classes, where distribution $Z_3$ was applied. It seems to be natural, since the more clusters appear in a configuration, the larger long-range connectivity is possible, bringing the sample closer to the so-called {\it percolation threshold}. Fig.~\ref{fig:EC_color} also demonstrates that the way the inclusions are distributed  (rather than the diversity of their radii) has much impact on the EC. 

\begin{figure}[ht]
\centering
\includegraphics[clip, trim=2mm 6mm 10mm 6mm, width=.75\textwidth]{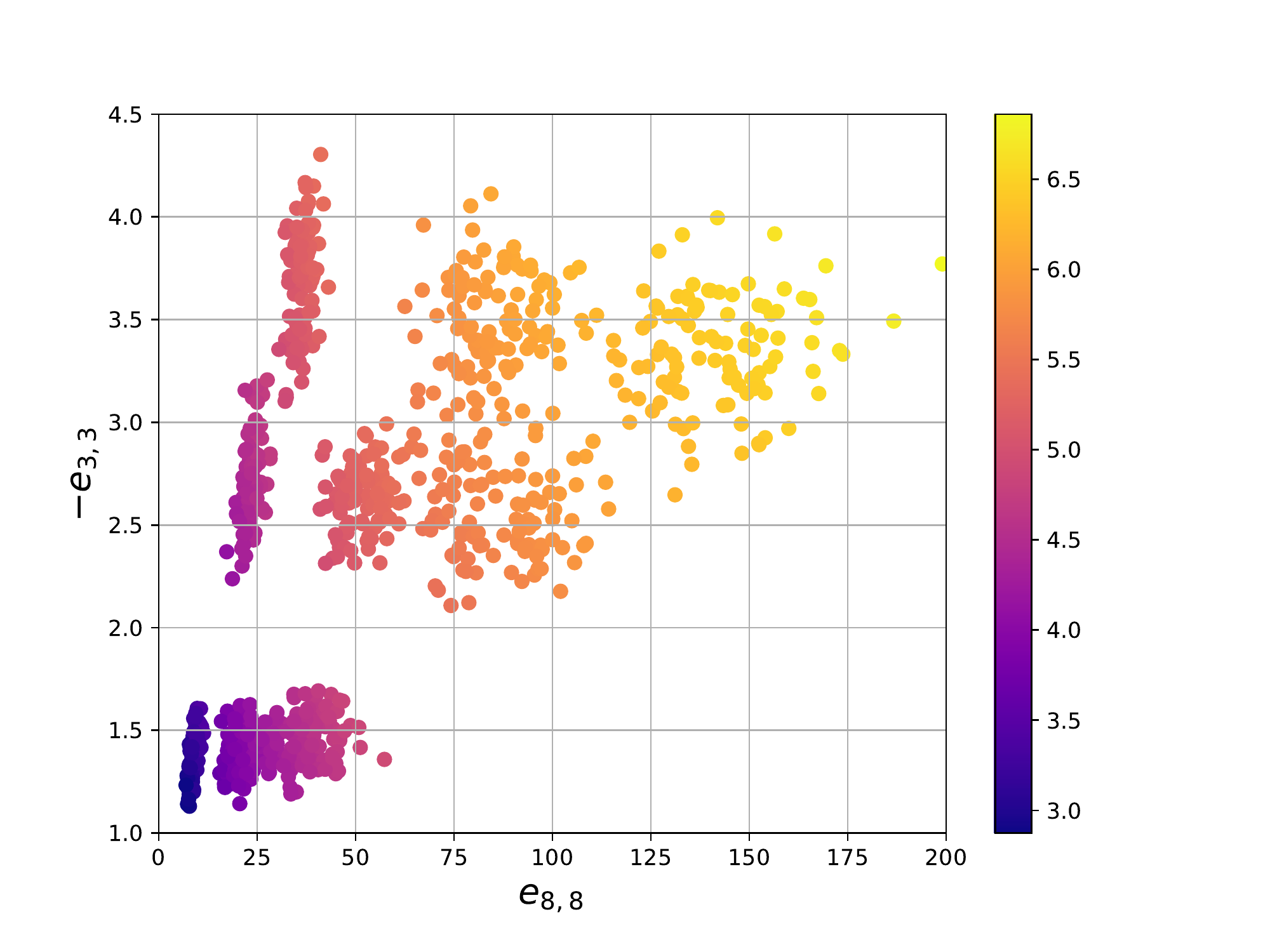} 
\caption{Values of $-e_{3,3}$ against $e_{8,8}$ for samples from from considered distributions. The colors represent the values of the irregularity measure for a given sample.} 
\label{fig:reg_color}
\end{figure}

Both selected sums can be combined into one {\it irregularity measure} of sample $s$ as follows:
\begin{equation}
\mu(s):= \log\left[ \left({1-e_{3,3}(s)}\right) \left({1+e_{8,8}(s)}\right)\right].
\end{equation}
Note that irregularity of the hexagonal array of identical disks is equal 0. The larger the value is, the more irregular model we expect (see Fig.~\ref{fig:reg_color}).
Table~\ref{tab:regularity} shows the mean values of the irregularity measure for each class of distributions. Note that the results are consistent with our initial guess.
\begin{table}[!htb]
\caption{Mean values of $\mu(s)$ computed for each considered class of distributions of disks.}
\label{tab:regularity}
\setlength{\tabcolsep}{3pt}
\centering
\begin{tabular}{lcclc}
\hline
 distribution & $\langle\mu(s)\rangle$ && distribution & $\langle\mu(s)\rangle$\\
\hline
hexagonal array                  & 0.000 && $Z_3$, identical radii           & 5.163\\
square array                     & 2.950 && $Z_1$, normally distr. radii     & 5.271\\
$Z_2$, identical radii           & 3.164 && $Z_1$, uniformly distr. radii    & 5.743\\
$Z_2$, normally distr. radii     & 4.003 && $Z_3$, normally distr. radii     & 5.957\\
$Z_1$, identical radii           & 4.505 && $Z_3$, uniformly distr. radii    & 6.429\\
$Z_2$, uniformly distr. radii    & 4.565 && &\\

\end{tabular}
\end{table}

\section{Conclusions}

In the present paper we defined the general form of the structural sums feature vector of composites with inclusions represented by distributions of non-overlapping disks. All considered examples clearly prove that the vector of structural sums carries large amount of information, for example, the inclusions' distribution protocol and the type of distributed inclusions. 

The paper also answers an important question, whether the higher-order sums are worth analysing or all the information is encoded in the lower-order parameters? We observed that the increase of training data has little effect on the performance of lower-order sums. Hence, in order to gain high accuracy, it was necessary to increase the order of the feature vector. Moreover, it was shown that different kind of data may require the application of a specific modification of features. 
 All investigated patterns were rather isotropic, hence the analysis of anisotropic models may reveal more details about structural sums.

As an application we selected two features and described their geometric meaning. Then, using selected sums, we introduced the irregularity measure of random structures. 
We also performed analysis and visualisation of the relationship between the effective conductivity of composites and the geometry of inclusions.

In our examples we used samples of 256 disks, however it is possible to compute structural sums for much larger systems ($\sim10^3$).
Since structural sums may seem to be computationally and conceptually complex, we are currently working on a software package providing high level of abstraction in calculations of the structural sums feature vector and plan to deploy it in the near future. 

We believe that there is a potential in structural sums as a tool of data analysis. The presented approach can be applied to other types of data as long as the data can be represented  by distributions of non-overlapping disks on the plane. The area of potential applications involves, for instance, plane objects frequently studied in biological and medical images. Arrangements of objects on the plane are also common in the study of self-assembly processes reflecting in individual components of different patterns \cite{Whitesides2002}. One can also consider a set of centres of disks as a point process pattern, hence the structural sums can be applied as characteristics of point fields, an important area of statistics \cite{ChiuBook}. In particular, disks with zero radii correspond to the Poisson process.  Such an approach is under development and will be published in a future paper.




\appendix

\section{Lattice sums and Eisenstein elliptic functions}
\label{eisenFun}
\renewcommand\thesection{A}

Consider the lattice~$\mathcal{Q}$ (see section~\ref{sect:theory}). For definiteness, it is assumed that $\textrm{Im}\;\tau>0$, where $\tau =\omega _{2}/\omega _{1}$. The Eisenstein summation is defined by the iterative sum
\begin{equation}
\sum_{m_{1},m_{2}}=\lim_{N\rightarrow \infty
}\sum_{m_{2}=-N}^{N}\left( \lim_{M\rightarrow \infty
}\sum_{m_{1}=-M}^{M}\right) .  \label{2.2}
\end{equation}
The lattice sums are introduced as follows
\begin{equation}
S_{n}:=\sum_{m_{1},m_{2}}\;^{\prime
}(m_{1}\omega _{1}+m_{2}\omega _{2})^{-n} \quad (n=2,3,\ldots), 
\label{2.1}
\end{equation}
where  the prime means that $m_{1}$ and $m_{2}$ run over all integer numbers  as in \eqref{2.2} except the pair $(m_{1},m_{2})=(0,0)$. The sum $S_{2}$ is conditionally convergent and understood in the sense of the Eisenstein summation (\ref{2.2}).  Though the rest sums (\ref{2.1}) converge absolutely, the direct computations by (\ref{2.1}) are problematic because of their slow convergence.  The sum $S_{2}$ can be computed by a quick formula \cite{ryl}:
\begin{equation}
S_{2}=
\left( \frac{\pi }{\omega
_{1}}\right)
^{2}\left( \frac{1}{3}-8\sum_{m=1}^{\infty }\frac{mq^{2m}}{1-q^{2m}}\right) ,%
\text{ where }q=\exp \left( \pi i\tau \right).  \label{2.3}
\end{equation}
It is known that $S_{n}=0$ for an odd $n$. For an even $n$, the sums (\ref{2.1}) can be easily
computed through the rapidly convergent infinite sums~\cite{MRP}
\begin{eqnarray}
S_{4}=\frac{1}{60}\left( \frac{\pi }{\omega
_{1}}\right)
^{4}\left( \frac{4}{3}+320\sum_{m=1}^{\infty }\frac{m^{3}q^{2m}}{1-q^{2m}}%
\right) ,\;  \label{2.5} \\
S_{6}=\frac{1}{140} \left( \frac{\pi }{\omega
_{1}}\right)
^{6}\left( \frac{8}{27}-\frac{448}{3}\sum_{m=1}^{\infty }\frac{m^{5}q^{2m}}{%
1-q^{2m}}\right).
\end{eqnarray}
The sums $S_{2n}$ ($n\geq 4$) are calculated by the recurrence formula~\cite{MRP}
\begin{equation}
S_{2n}=\frac{3\sum_{m=2}^{n-2}\left( 2m-1\right)
\left( 2n-2m-1\right) S_{2m}S_{2(n-m)}}{\left( 2n+1\right) \left(
2n-1\right) \left( n-3\right) }.  \label{2.7}
\end{equation}

\noindent The Eisenstein series are defined as follows~\cite{weil}
\begin{equation}
E_{n}(z):=\sum_{m_{1},m_{2}}(z-m_{1}\omega
_{1}-m_{2}\omega _{2})^{-n}\,,\;n=2,3,...\;.  \label{2.8}
\end{equation}
Each of the functions (\ref{2.8}) is doubly periodic and has a pole of order
$n$ at $z=0$. Further, it is convenient to define the
value of $E_{n}(z)$ at the origin as $E_{n}(0):=S_{n}.$

The Eisenstein functions $E_n(z)$  ($n=2,3,\ldots$) and the Weierstrass function $\wp(z)$ are related by the identities~\cite{MRP,weil}
\begin{equation}
	\begin{array}{cc}
	E_{2}(z)=\wp(z)+S_2,\\ \\
	E_{n}(z)=\frac{(-1)^n}{(n-1)!}\frac{d^{n-2}}{dz^{n-2}}\wp(z), \quad n=3, 4\ldots	\\
	\end{array}
\label{eq:EtoP1}
\end{equation}
where $z\neq 0$ and $S_2$ is a constant.
It follows from the elliptic function theory~\cite{Akhiezer1} that
\begin{equation}
\wp''(z)=6\wp(z)^2-30 S_4.
\label{eq:wpBis}
\end{equation}


\end{document}